\documentclass{article}

\usepackage{arxiv}
\usepackage{xcolor}
\usepackage[utf8]{inputenc} 
\usepackage[T1]{fontenc}    
\usepackage{hyperref}       
\usepackage{url}            
\usepackage{booktabs}       
\usepackage{amsfonts}       
\usepackage{nicefrac}       
\usepackage{microtype}      
\usepackage{lipsum}		
\usepackage{paracol}
\usepackage{lineno}
\usepackage{graphicx}
\usepackage{amssymb}
\usepackage{caption}
\DeclareCaptionLabelFormat{supp}{Supp. Fig. #2}
\usepackage{changepage}
\usepackage{natbib}
\usepackage{doi}
\usepackage{amsmath}
\DeclareMathOperator*{\argmax}{arg\,max}

\title{Controlling Biofilm Transport with Porous Metamaterials Designed with Bayesian Learning}

\author{ Hanfeng Zhai, \ \ Jingjie Yeo\\
	Sibley School of Mechanical and Aerospace Engineering\\
	Cornell University, Ithaca, NY 14850, USA}

\renewcommand{\shorttitle}{Bayesian Optimization for Bioporous Materials Design}
\DeclareCaptionLabelFormat{mysuplabel}{Supp. Fig. \arabic{figure}}

\begin{document}


\maketitle

\begin{abstract}
Biofilm growth and transport in confined systems frequently occur in natural and engineered systems. Designing customizable engineered porous materials for controllable biofilm transportation properties could significantly improve the rapid utilization of biofilms as engineered living materials for applications in pollution alleviation, material self-healing, energy production, and many more. We combine Bayesian optimization (BO) and individual-based modeling to conduct design optimizations for maximizing different porous materials’ (PM) biofilm transportation capability. We first characterize the acquisition function in BO for designing 2-dimensional porous membranes. We use the expected improvement acquisition function for designing lattice metamaterials (LM) and 3-dimensional porous media (3DPM). We find that BO is 92.89\% more efficient than the uniform grid search method for LM and 223.04\% more efficient for 3DPM. For all three types of structures, the selected characterization simulation tests are in good agreement with the design spaces approximated with Gaussian process regression. All the extracted optimal designs exhibit better biofilm growth and transportability than unconfined space without substrates. Our comparison study shows that PM stimulates biofilm growth by taking up volumetric space and pushing biofilms’ upward growth, as evidenced by a 20\% increase in bacteria cell numbers in unconfined space compared to porous materials, and 128\% more bacteria cells in the target growth region for PM-induced biofilm growth compared with unconfined growth. Our work provides deeper insights into the design of substrates to tune biofilm growth, analyzing the optimization process and characterizing the design space, and understanding biophysical mechanisms governing the growth of biofilms.
\end{abstract}



\section{Introduction\label{sec_intro}}

Biofilms, commonly defined as surface-attached communities of microorganisms (i.e., groups of bacteria cells) embedded in a self-produced matrix of extracellular polymeric substances (EPS) \cite{biofilmmedref1}, grow mostly in confined systems such as rock cracks, industrial pipelines, biological bodies, and many other artificial or natural microenvironments \cite{pnas_activesurface}. One of the prerequisites of biofilm growth is the existence of adhesive surfaces that allow bacteria to grow and cluster into ``film-shaped'' communities, aided by adhesive EPS. Hence, increasing surface areas would allow biofilm observed in mostly confined systems to attach and grow further \cite{surface_area_biofilm}. From the engineering perspective, biofilms possess abundant pros and cons to human society. On the negative side, the formation and attachment of biofilms pose serious problems for marine engineering by fouling the surfaces of marine vessels, equipment, and infrastructure, leading to reduced efficiency and increased maintenance costs \cite{marineref1, marineref2}. On biomedical devices, such as catheters and implants, biofilm formation can lead to infections that are difficult to treat \cite{biofilmmedref1, biofilmmedref2}. On the positive side, biofilm can also be utilized as {\em engineered living materials} (ELM) with important engineering applications such as for creating self-healing concrete by incorporating bacteria into the concrete mix \cite{concreteref1}, for treating wastewater by removing pollutants and nutrients \cite{wastewatertreatref}, and for 3D bioprinting into functional soft materials \cite{3dbioprintingref}.

Considering all these pros and cons, understanding the mechanisms of biofilm growth within confined systems is crucial for humanly-desirable control of biofilm, particularly in three major applications: (1) Prevent undesired biofilm attachment and conduct efficient biofilm removal \cite{zhai_acs_biofilm}. (2) Using biophysics to promote the effective usage of biofilm as ELM, e.g., clean energy applications \cite{biofilm_clean_energy}. (3) By combining both the pros and cons to enable biofilm control to design customized devices and sensors \cite{biofilm_control_ref}. Towards achieving such applications, we identify a major design challenge for biofilm control: designing porous structural materials that can control biofilm growth. However, two problems naturally arise in our road towards efficient biofilm control and utilization: (1) Conducting experiments on biofilm is time-consuming due to the lengthy growth process, hence characterizing and benchmarking changes in the bacterial environment cannot be achieved in a time-efficient manner; (2) Directly modifying the structures of porous materials in a multi-parametric way to test the corresponding biofilm growth is not straightforward and causes the investigation to be even more time-consuming. Hence, novel techniques that can bypass this ``trial-and-error'' approach is of urgent need. 


To tackle the problem of time-consuming experiments, we use computational modeling, more specifically, individual-based modeling. Notably, there are various computational modeling methods have been proposed in recent years to model biofilm. For example, molecular dynamics simulations were used to model the biochemical properties of biofilm on the molecular scale \cite{md_biofilm_study}, dissipative particle dynamics were used to model biofilm deformation under shear flow \cite{dpd_biofilm_model}, coarse-grained molecular dynamics were used to study dewetting phenomena \cite{cgmd_biofilm}, and finite element methods were used to simulate the linearized growth \cite{biofilm_fem_model} to model biofilm on the continuum scale. Here, we use individual-based modeling (IbM) \cite{Li2019} that represents each bacteria cell as individual spherical particles in combination with mathematical models for the growth and dynamics of biofilm. IbM captures the behavior of biofilms in length scales that range from individual bacterial cells to clusters of cells while requiring relatively low computational resources. Most importantly, we are motivated by three critical considerations:\begin{itemize}
    \item IbM is a general multiscale method, capable of capturing the scaling effects from cell to ``film''. Since each bacteria cell is modeled as a discrete particle, interactions between ``cell-cell''\footnote{multiple cells constitute the biofilm, while ``cell-cell'' interactions dictate the dynamics of individuals within the biofilms}, ``biofilm-material'', and ``cell-material'' can be also be modeled correspondingly. When studying the transport of biofilm within porous regions, the ability to capture such multiscale mechanics is essential as both the individual and group dynamics play important roles. \cite{Li2019}.
    \item The IbM method is physically realistic for the spatiotemporal scale of interest, where IbM can capture the dynamics and mechanics of the biofilms observed in natural pores with diameters on the order of $10^{-5}\sim10^{-3}$m \cite{biofilm_porescale_mircometer}. Each bacteria cell is approximately 1$\mu$m, hence our modeled porous structures perfectly capture the local morphologies of the biofilms as they grow and propagate. Moreover, the physical adhesion and other micromechanisms that govern the overall mechanical behavior of the biofilms primarily originate from the micrometer scale \cite{biofilm_mechanical_adhesion_micrometer}, which can be directly observed and quantified using IbM. Also, our ultimate goal is to incorporate our theoretical predictions and understanding into experimental designs of ELM and the most recent work on ELM is on the micrometer scale \cite{englivmat_natrevmat_reviewpaper} which is directly relevant to our IbM.
    \item Compared with other methods, IbM has the most decent computational burden requirement for relatively high fidelity\footnote{The goal is to combine simulation with optimization, where the simulation is treated as the evaluated function. Hence, the function evaluation time is important for efficient optimizations.}. Simulating the growth of biofilm at the molecular scale using methods like molecular dynamics (MD) simulation or Monte Carlo sampling will require impossibly large amounts of computational resources and time. As a reference, 6 months of time is needed to run MD simulations of a large protein structure for 1 $\mu$s \cite{md_fulltime_6months}, making this method infeasible for our problem. On the flip side, at the continuum scale, simulations of biofilm usually incorporate extended finite element methods (XFEM) and level set method (LSM) \cite{biofilm_xfem_levelset_ijnme}, which can be extremely computationally burdensome as these methods usually require a moving mesh that resolves the phase boundary \cite{bubblenet}, which is computationally costly while being unable to capture the dynamics of individual cells.
\end{itemize}

To solve the problem of inefficient forward predictions of structural designs, we use approximation methods to solve the inverse problem of materials design. Suppose one were to define designing materials by perturbing their original structures to obtain the target properties as a {\em forward problem}. In that case, one can then define obtaining the tailored materials' structures from the predefined targeted properties as an {\em inverse problem}\footnote{The rigorous formulation follows the {\em Hadamard's principles}, which we do not discuss in details here.} The detailed inverse problem here is formulated as finding the optimal porous structure corresponding to the target biofilm transport properties (i.e., maximizing biofilm growth), as a class 2 inverse problem. However, there are two main specifics to note: (A) The defined inverse problem is ill-posed \cite{hadamard_illposed}. Two or more different porous structures may yield the same biofilm transport properties, such that the material's structure found as the solution to the inverse problem may not be unique. (B) There are no analytical (or symbolic) forms of the inverse map. The biofilm simulation is constituted of iterative growth and update of bacteria cells, where it is almost impossible to obtain an analytical inverse of this coupled multiphysics system with changed parameters\footnote{For the detail of the simulation algorithms please refer to Refs. \cite{Li2019} \& \cite{zhai_acs_biofilm}}. 

To solve the ill-posed problem (A), we characterize the design space to approximate a surrogate model of the design space. We verify the approximated map by conducting verification simulations along the observed maximal solution and randomly selected points. This allows us to verify the accuracy of the fitted surrogate and that further analyses are reliable. To solve the problem of the lack of analytical forms of the inverse map (B), we avoid gradient-based optimizations and use machine learning (ML) techniques (specifically, Gaussian process regression (GPR)) for direct surrogate modeling of the design space, while also enabling our ability to characterize this design space. Hence, to implement the two solutions we proposed, we perform Bayesian optimization (BO) \cite{bayes}, using GPR to approximate the design space map and an acquisition function to update the solution search scheme. There are three major reasons for choosing BO:\begin{itemize}
    \item The flexibility of handling complex problems. Compared with gradient-based methods, BO is flexible and can be adapted to solving complicated optimization problems without requiring the calculation of the derivative of the evaluated functions.
    \item It is less computationally burdensome compared with other ML methods. As a non-parametric method, GPR requires less computational resources compared with neural networks (NN) and is especially suitable for problems defined within the limited data regime \cite{Jan_PIGPR}. Compared with the widely used deep reinforcement learning (DRL) \cite{drl_intro_book_ref}, BO does not require iterative training of the deep NN for each function evaluation, and hence is significantly less computationally burdensome.
    \item Approximating the design space map allows direct characterization and analysis of the sampling process. In metaheuristic methods such as genetic algorithms \cite{ga_paper_ref} or particle swarm optimization \cite{pso_paper_ref}, the function evaluations are based on random perturbations of the input variables inspired by natural phenomena. In contrast, the learned design space mapped using GPR can be characterized in detail. Moreover, BO usually does not heavily rely on data population while only requiring one evaluation per iteration, thus speeding up the characterization.
\end{itemize}

In this study, we combine IbM and BO to solve a focused problem: Inversely design the porous structural materials for biofilm transport and characterize the biomechanics from the optimization processes. By solving this problem, we aim to answer the following questions: (1) What are the optimal porous microstructures that can maximize the transport of biofilms? (2) Are the approximated design space accurate and how do we verify them? (3) What biomechanical mechanisms are discovered by optimizing and characterizing the design space?

In Section \ref{sec_method}, we briefly introduce the methods used, including our computational models of biofilm physics (Section \ref{subsec_compmod}), the BO scheme (Section \ref{subsec_bo}), such as surrogate modeling with GPR (Section \ref{subsubsec_gpr}) and the iterative update scheme using an acquisition function (Section \ref{subsubsec_acqfunc}), followed by three numerical experiments on different porous materials in Section \ref{subsec_numexp}. These results are discussed in Section \ref{sec_resdis}: the optimization processes and optimal structures for the different numerical experiments in Sections \ref{subsec_2dmat} \& \ref{subsec_3dmat}, verification of the discovered new phenomena that certain porous structures stimulate the growth of biofilms, and additional mechanistic explanations in Section \ref{subsec_biomechanics}. We then conclude our studies in Section \ref{sec_conclusion}.

\section{Methods\label{sec_method}}

As elaborated in Section \ref{sec_intro}, we will use IbM to model the growth of biofilms and their mechanical interactions with the porous metamaterials in a predefined simulation box. {\color{black}Here, the term ``metamaterials'' stand for mechanically architectured scaffolds that are employed to control the biomass transport at the ``film-scale''.} We then combine the BO methods together with the material representation of the porous structure parameterized based on our defined numerical experiments and the simulation framework to iteratively search for optimal porous structures that enhance biofilm transport properties. 

\begin{figure}[htbp]
    \centering
    \includegraphics[width=45em]{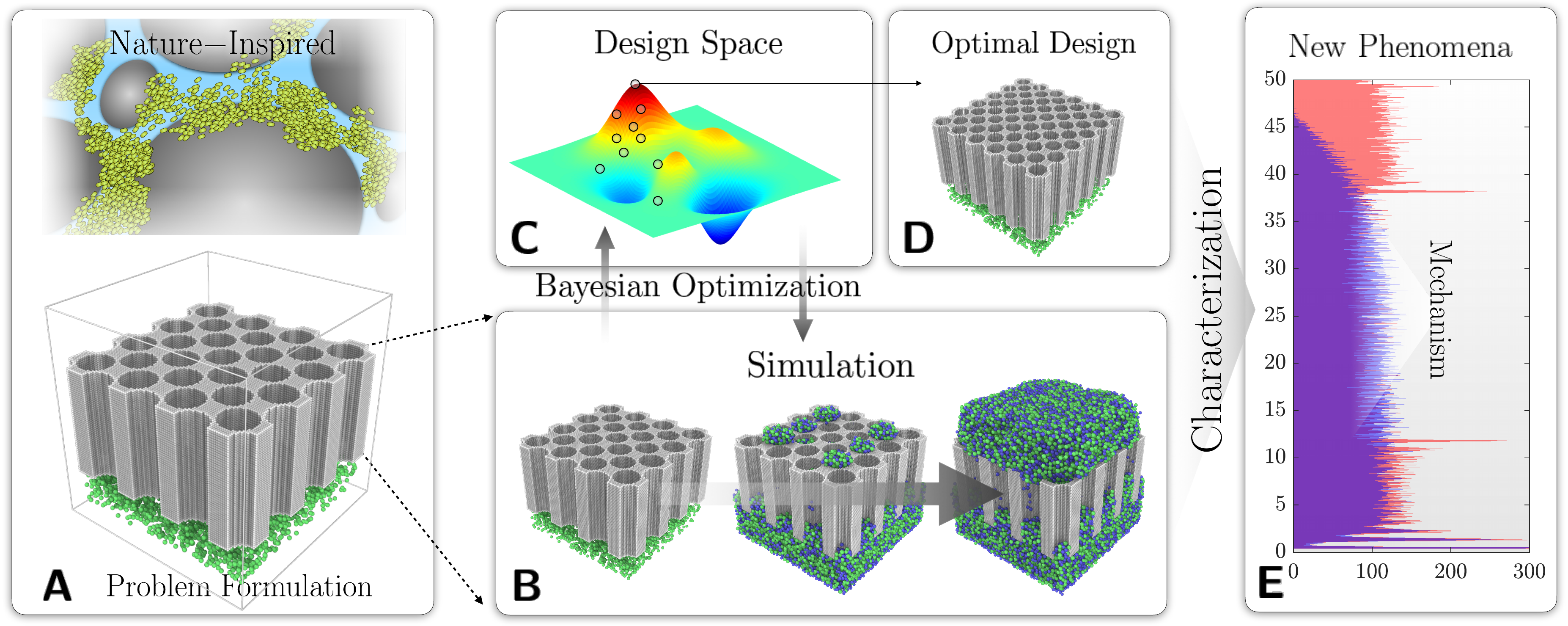}
    \caption{The overall schematic of this study. \textbf{(A)} Inspired by biofilm transport in porous materials (top), we define a computational framework (bottom), where bacteria cells seeded at the bottom grow into the porous substrates as indicated by the grey area. \textbf{(B)} The growth processes of the biofilms within the porous materials are simulated using IbM. \textbf{(C)} BO is then used to reconstruct the design space for the porous material and \textbf{(D)} the extracted optimal design is extracted from the design space. \textbf{(E)} New physical phenomena and the mechanism of bacteria transport in porous materials are uncovered by comparing the optimal design(s) against benchmark cases. }
    \label{fig1}
\end{figure}

The general schematic of this study is represented in Figure \ref{fig1}. We are inspired by a natural phenomenon: biofilms mostly grow in confined systems \cite{pnas_activesurface}, hence we construct porous structures that allow the biofilm to grow within to mimic this phenomenon (Figure \ref{fig1} \textbf{A}). To determine the optimal porous structures for biofilm growth, we then run the simulations initiated by parameterized materials representation (Figure \ref{fig1} \textbf{B}) for coupling with BO (Figure \ref{fig1} \textbf{C}). The coupling is enabled by ``variable passing'' between the simulation and optimization: the simulation takes the materials' representation as input and outputs the biofilm transport properties as the objective for optimization; the optimization algorithm then updates and outputs the new materials' representation for an iterative loop. This iterative search will eventually propose an optimal design (Figure \ref{fig1} \textbf{D}). By characterizing the design space obtained in the optimization (Figure \ref{fig1} \textbf{C}) and comparing the observation from these simulations, we then propose explanations for the optimal structures and identify new mechanisms of biofilm transport physics (Figure \ref{fig1} \textbf{E}).

In the following subsections, we first briefly introduce the basic formulation of our computational methods of IbM and the basic mathematical formulation of BO. We also briefly introduce the different numerical experiments of biofilm growth for porous membranes, lattice metamaterials, and nonconvex porous media, respectively.

\subsection{Computational Models\label{subsec_compmod}}

In this work, we used IbM based on the Newcastle University Frontiers in Engineering Biology (NUFEB) framework \cite{Li2019}, in which each bacteria cell is modeled as a spherical particle. Biofilms are formed by cell division and extrusion of EPS. Following our previous study on surface topology optimization \cite{zhai_acs_biofilm}, 
the following differential equation governs the microbe growth and decay:

\begin{equation}
    \frac{d m_i}{dt} = \xi_i m_i \label{equation_growth}
\end{equation}

where $m_i$ is the biomass of the $i^{th}$ bacteria cells and $\xi_i$ is the growth rate. The growth rate of each bacteria cell is $\xi = 0.00028s^{-1}$. To avoid particles overlapping while growing, which will de-stabilize the simulations, the particles are mechanically relaxed according to Newton's equation 

\begin{equation}
    m_i \frac{d{\bf v}_i}{dt} = \mathbf{F}_{c,i} + \mathbf{F}_{a,i} \label{newton_grow}
\end{equation}

where $\mathbf{v}_i$ is the particles' velocity. The contact force $\mathbf{F}_{c,i}$ is a pair-wise force between particles to prevent overlapping based on Hooke's law 

\begin{equation}
    \mathbf{F}_{c,i} = \sum_{j=1}^{N_i} \left( K_\mathbb{N} \delta \mathbf{n}_{i,j} - m_{i,j} \gamma_\mathbb{N}  \mathbf{v}_{i,j} \right)\label{f1}
\end{equation}

where $N_i$ is the total number of neighboring particles of $i$, $K_\mathbb{N}$ is the elastic constant for normal contact, $\delta \mathbf{n}_{ij}$ is the overlap distance between the center of particle $i$ and its neighbor particle
$j$. $\gamma_\mathbb{N} $ is the viscoelastic damping constant for normal contact, and $v_{i,j}$ is the relative velocity of the two particles. The EPS adhesive force ${\bf F}_{a,i}$ is a pair-wise interaction modeled as a van der Waals force

\begin{equation}
    \mathbf{F}_{a,i} = \sum_{j=1}^{N_i} \frac{H_a r_{i,j}}{12 h_{min,i,j}^2 } \mathbf{n}_{i,j}\label{f2}
\end{equation}

where $H_a$ is the Hamaker coefficient, $r_{i,j}$ is the effective outer-radius of the $i^{\rm th}$ and $j^{\rm th}$ particles. $h_{min,i,j}$ is the minimum separation distance of the two particles, and $\mathbf{n}_{i,j}$ is the unit vector from particle $i$ to $j$. 

Mechanical equilibrium is achieved when the average pressure of the microbial community reaches a plateau. The average pressure $P$ of the system is calculated as

\begin{equation}
    P = \frac{1}{3V} \left( \sum_{i=1}^N m_i \mathbf{v}_i \cdot \mathbf{v}_i + \sum_{i=1}^N\sum_{j>i}^N \mathbf{r}_{i,j} \cdot\mathbf{F}_{i,j} \right)\label{relax_equilibrium_mechanical}
\end{equation}

where $V$ is the sum of the particles' volumes. The first term in the bracket is the contribution from the kinetic energy of each particle. The second term is the interaction energy, where $\mathbf{r}_{i,j}$ and $\mathbf{F}_{i,j}$ are the distance and force between two interacting particles $i$ and $j$, respectively.

Here, the bacterial growth rate is determined by the Monod kinetic equation \cite{Monod1949} driven by the local concentration of nutrients. The porous substrates are modeled as fully rigid particles with neither growth nor decay. Here, under the Monod model formulation, each bacteria cell first grows with increasing radii, and after their radii reach a critical value $r^{\sf C}= 1.36\times10^{-6}$m, the cell is separated into two daughter cells (full details are in Ref. \cite{Li2019}). The EPS, also modeled as spherical particles, are secreted by the main bacteria cells in the growing process (full details are in Refs. \cite{eps_original_formulation, nufeb_eps_formulation}). After a pre-defined number of iterations, the system will arrive at a total number of bacteria cells and the EPS particles, which we denote as $\mathcal{N}_{\rm bio}^{\rm total}$.

{\color{black}The model we implemented assumes a constant growth rate under linear nutrient gradients\footnote{Further discussions can be checked in Section \ref{subsec_limitations}}. Qualitatively, to ensure that the IbM reproduces experimentally observed biofilm behavior, a few parameters are of importance: (1) The growth rate $\xi_i$ should match the experimentally observed value. The actual biofilm growth dynamics should also not strongly deviate from a constant growth rate. (2) The modelled bacteria should be a Heterotroph (HET), as the IbM uses the Monod kinetics growth model for Heterotrophic bacteria. If there are additional bacteria types, more complex IbM models should be considered \cite{Li2019, nufeb_eps_formulation}. (3) Additional model parameters, such as nutrient distribution and diffusion coefficients, should match the environment of the bacteria (Supp. Tab. 1). (4) The viscosity damping constant $\gamma_\mathbb{N}$ should match with viscosity tested based on bacteria surround fluids. (5) The normal contact constant $K_\mathbb{N}$ should also match with the parameter tested based on the bacteria cell's mechanical properties. }


\subsection{Bayesian Optimization\label{subsec_bo}}


The goal of optimization is to minimize or maximize an objective function, which in our case is the bacteria cell number under a target design region, denoted as $\mathcal{N}_{\rm bio}$ for ease of notation ($\mathcal{N}_{\rm bio}\subset\mathcal{N}_{\rm bio}^{\rm total}\in \mathbb{Z}$). Using $\mathcal{N}_{\rm bio} = \mathcal{M}_{\sf NUFEB}(N_{\rm unit}, \bar{\mathcal{D}}; \mathbf{p})$ to denote a multivariate function relation, in which $N_{\rm unit}$ and $\bar{\mathcal{D}}$ stand for unit cell numbers per simulation box side and the dimensionless structural parameter (or dimensionless variable), respectively. $N_{\rm unit}$ and $\bar{\mathcal{D}}$ are the design variables and further details are in Section \ref{subsec_numexp}. For simplicity, we use $\mathcal{DV} = [N_{\rm unit}$, $\bar{\mathcal{D}}]$ to denote the design variables. $\bf p$ are the parameters used in the IbM simulations, as presented in Equations (\ref{equation_growth}$\sim$\ref{relax_equilibrium_mechanical}). The optimization process can be simplified as: 

\begin{equation}
    \begin{aligned}
    \argmax_{N_{\rm unit}, \bar{\mathcal{D}}}\mathcal{N}_{\rm bio} = {\mathcal{M}_{\sf NUFEB}}(N_{\rm unit}, \bar{\mathcal{D}}; \mathbf{p}),\\
    {\rm subject\ to}\quad { \bar{\mathcal{D}}}_{\rm LB} \leq \bar{\mathcal{D}} \leq \bar{\mathcal{D}}_{\rm UB},\ 1 \leq N_{\rm unit}\leq 15\ ( N_{\rm unit}\in \mathbb{Z})
    \end{aligned}\label{opt_prob_equation}
\end{equation}

Here, we define a target growth region to count $\mathcal{N}_{\rm bio}$ (Section \ref{subsec_numexp}), so that the materials' microstructure will be optimized to enhance growth towards the targeted region. Given the input design variables $\mathcal{DV}$, we represent the biofilm physics growth simulation model as a map, $\mathcal{M}_{\sf NUFEB}: N_{\rm unit}, \bar{\mathcal{D}} \rightarrow \mathcal{N}_{\rm bio}$, where the simulation parameters $\mathbf{p} = [\xi_i, K_\mathbb{N}, \gamma_\mathbb{N}, H_a, r^\mathsf{C},...]$ are incorporated in the IbM model (Section \ref{subsec_compmod}). $\mathcal{M}_{\sf NUFEB}(\cdot)$ denotes IbM simulations that map the design representation of the materials as input and the bacterial cell number count as output. $N_{\rm unit}$ is an integer between 1 and 15 as the number of unit cells are changing during the BO iterations. The dimensionless structure parameter $\bar{\mathcal{D}}$ is defined per case, as the lower and upper bounds, $\bar{\mathcal{D}}_{\rm LB}$ and $\bar{\mathcal{D}}_{\rm UB}$, differ based on the simulation and materials basis settings, to be discussed in Section \ref{subsec_numexp}.

BO aims to iteratively update new evaluations from the computational models in Section \ref{subsec_compmod} to search for optimal porous structures. By sampling multiple simulations and mapping the design variables onto the defined objective, we construct a surrogate of the direct map between the input (i.e., the design variables) and the output (i.e., the objective) from GPR. This GPR reconstructed surrogate is then updated through the acquisition functions of choice.


\subsubsection{Gaussian Process Regression\label{subsubsec_gpr}}



GPR is a Bayesian statistical approach to approximate and model function(s). Considering our optimization problem, the function can be denoted as $\mathcal{N}_{\rm bio} = \mathcal{M}_{\sf NUFEB}(\mathcal{DV};\mathbf{p})$, where $\mathcal{N}_{\rm bio}$ is evaluated at a collection of different sets of points (or design variables): $\mathcal{DV}_1, \mathcal{DV}_2, ..., \mathcal{DV}_k \in \mathbb{R}^2$. We can obtain the vector $[\mathcal{M}_{\sf NUFEB}(\mathcal{DV}_1), ..., \mathcal{M}_{\sf NUFEB}(\mathcal{DV}_k)]$ to construct a surrogate model for the design parameters with the correlated objectives. The vector is randomly drawn from a prior probability distribution, where GPR takes this prior distribution to be a multivariate normal with a particular mean vector and covariance matrix. Here, the mean vector and covariance matrix are constructed by evaluating the mean function $\mu_0$ and the covariance function $\Sigma_0$ at each pair of points $\mathcal{DV}_i$, $\mathcal{DV}_j$. The resulting prior distribution on the vector $[\mathcal{M}_{\sf NUFEB}(x_1),..., \mathcal{M}_{\sf NUFEB}(x_k)]$ is represented in the form of a normal distribution to construct the surrogate model \cite{bayes}:

\begin{equation}
    \mathcal{N}_{\rm bio}(\mathcal{DV}_{1:k}) \sim {\mathfrak{N}}\left(\mu_0 (\mathcal{DV}_{1:k}), \Sigma_0 (\mathcal{DV}_{1:k}, \mathcal{DV}_{1:k}))\right)\label{surrogate}
\end{equation}

where $\mathfrak{N}(\cdot)$ denotes the normal distribution. The collection of input points is represented in compact notation: $1:k$ represents the range of $1,2,..., k$.

The surrogate model $\mathcal{M}_{\sf NUFEB}(\mathcal{DV})$ on $1:k$ is represented as a probability distribution given in Equation (\ref{surrogate}). To update the model with new observations, such as after inferring the value of $\mathcal{M}_{\sf NUFEB}(\mathcal{DV})$ at a new point $\mathcal{DV}$, we let $k = l+1$ and $\mathcal{DV}_k = \mathcal{DV}$. The conditional distribution of $\mathcal{N}_{\rm bio}$ given observations $\mathcal{DV}_{1:l}$ using Bayes' rule is

\begin{equation}
    \begin{aligned}
   \mathcal{N}_{\rm bio}(\mathcal{DV})| \mathcal{N}_{\rm bio}(\mathcal{DV}_{1:l}) &\sim \mathfrak{N}(\mu_l (\mathcal{DV}), \sigma_l^2 (\mathcal{DV}))\\
    \mu_l (\mathcal{DV}) &= \Sigma_0 (\mathcal{DV}, \mathcal{DV}_{1:l}) \Sigma_0 (\mathcal{DV}_{1:l},\mathcal{DV}_{1:l})^{-1} \left(\mathcal{M}_{\sf NUFEB}(\mathcal{DV}_{1:l}) - \mu_0 (\mathcal{DV}_{1:l})+\mu_0(\mathcal{DV}) \right)\\
    \sigma_l^2 &= \Sigma_0 (\mathcal{DV}, \mathcal{DV}) - \Sigma_0 (\mathcal{DV}, \mathcal{DV}_{1:l})\Sigma_0 (\mathcal{DV}_{1:l}, \mathcal{DV}_{1:l})^{-1} \Sigma_0 (\mathcal{DV}_{1:l}, \mathcal{DV})
    \end{aligned}
\end{equation}

where the posterior mean $\mu_l(\mathcal{DV})$ is a weighted average between the prior $\mu_0(\mathcal{DV})$ and the estimation from $\mathcal{M}_{\sf NUFEB}(\mathcal{DV}_{1:l})$, where the weight applied depends on the Gaussian kernel. {\color{black}Here, we employ the Matérn kernel function to compute the covariance, with the positive parameter $\nu = 2.5$ \cite{gp_book}}. Our goal is to estimate the parameters $\sigma$ and $\theta_m$ that create the surrogate model given the training data $[(\mathcal{N}_{\rm bio})_k,\ \mathcal{DV}_k]$ at iteration $k$. Here, we will use $\hat{\mathcal{M}}_{\sf GPR}$ to denote the surrogate model constructed from GPR in the iterative updating process. The updating sampling scheme is achieved through the acquisition function in the following section, which improves the accuracy of the updated surrogate so that the reconstructed design space approximates the theoretical continuous design from NUFEB simulations $\hat{\mathcal{M}}_{\sf GPR}\sim \mathcal{M}_{\sf NUFEB}$.

\subsubsection{Acquisition Function\label{subsubsec_acqfunc}}

Given the training data $[(\mathcal{N}_{\rm bio})_k,\ \mathcal{DV}_k]$, Equation (\ref{surrogate}) gives us the prior distribution $(\mathcal{N}_{\rm bio})_l \sim \mathfrak{N}(\mu_0, \Sigma_0)$ as the surrogate. This prior and the given dataset induce a posterior: the acquisition function denoted as $\mathcal{A}: \mathcal{X}\xrightarrow{} \mathbb{R}^+$, determines the point in $\mathcal{X}$ to be evaluated through the proxy optimization $\mathcal{DV}_{\rm best} = \argmax_{\mathcal{DV}}\mathcal{A}(\mathcal{DV})$. The acquisition function depends on the previous observations, which can be represented as $\mathcal{A} = \mathcal{A}(\mathcal{DV}; (\mathcal{DV}_l, (\mathcal{N}_{\rm bio})_l), \theta)$, where $(\mathcal{DV}_l, (\mathcal{N}_{\rm bio})_l)$ leads to the reconstructed $\hat{\mathcal{M}}_{\sf GPR}$. Based on our mathematical notations, the new observation is probed through the acquisition function \cite{msde_nanoporous_material}:

\begin{equation}
    \mathcal{DV}_{k}=\mathcal{DV}_{l+1}= \argmax_{\mathcal{DV}} \mathcal{A}\left( \mathcal{DV};(\hat{\mathcal{M}}_{\sf GPR})_l, \theta_m\right)
\end{equation}

where the input space contains the evaluation of design variables at $l$ points: $(\mathcal{DV}_1, \mathcal{DV}_2, ..., \mathcal{DV}_l)$. We compare and characterize two different acquisition functions, the Upper Confidence Bound (UCB) and the Expected Improvement (EI), to benchmark the effect of acquisition updates. The UCB exploits the upper confidence bounds to construct the acquisition function and minimize regret. UCB takes the form \cite{acquisition_func} 

\begin{equation}
    \mathcal{A}_{\sf UCB}\left(\mathcal{DV}; (\mathcal{DV}_l, (\mathcal{N}_{\rm bio})_l), \theta_m\right):= \mu_l \left(\mathcal{DV}; (\mathcal{DV}_l, (\mathcal{N}_{\rm bio})_l), \theta_m\right) + \kappa \sigma\left(\mathcal{DV}; (\mathcal{DV}_l, (\mathcal{N}_{\rm bio})_l), \theta_m\right)\label{acquisition_function_EI}
\end{equation} 

where $\kappa$ is a tunable parameter balancing exploitation and exploration when constructing the surrogate model. We take $\kappa=2$ in our implementations. For the EI acquisition, the function writes:

\begin{equation}
    \begin{aligned}
        \mathcal{A}_{\sf EI} \left(\mathcal{DV}; (\mathcal{DV}_l, (\mathcal{N}_{\rm bio})_l), \theta_m\right):= \sigma_l \left(\mathcal{DV}; (\mathcal{DV}_l, (\mathcal{N}_{\rm bio})_l), \theta_m\right) \left(\gamma(\mathcal{DV}){\bf\Phi} \left(\gamma(\mathcal{DV})\right) + \mathfrak{N} \left(\gamma(\mathcal{DV}); 0,1\right)\right)\label{acquisition_function_UCB}
    \end{aligned}
\end{equation}

where $\gamma$ is computed as $\gamma = \left(-\mathcal{M}_{\sf NUFEB}(\mathcal{DV}_{\rm best}) + \mu(\mathcal{DV}; \{\mathcal{DV}_l, (\mathcal{N}_{\rm bio})_l\}_l, \theta) - \Xi\right)/ \sigma\left(\mathcal{DV}; \{\mathcal{DV}_l, (\mathcal{N}_{\rm bio})_l\}_l, \theta \right)$, where $\Xi$ is a damping factor in the code implementation, and $\Xi=10^{-4}$ in our implementation. Note that $\mathcal{A}_{\sf EI}$ preserves a closed form under the GP evaluations.

Combining GPR and the acquisition function, the surrogate model can approximate the design space's maximal value. In our case, such BO methods optimize porous structures to achieve maximal bacterial cell numbers in the targeted region. Here, the total function evaluations are different per case, as to be discussed in the following Section \ref{subsec_numexp}.

\subsection{Numerical Experiments\label{subsec_numexp}}

\begin{figure}[htbp]
    \centering
    \includegraphics[width=42em]{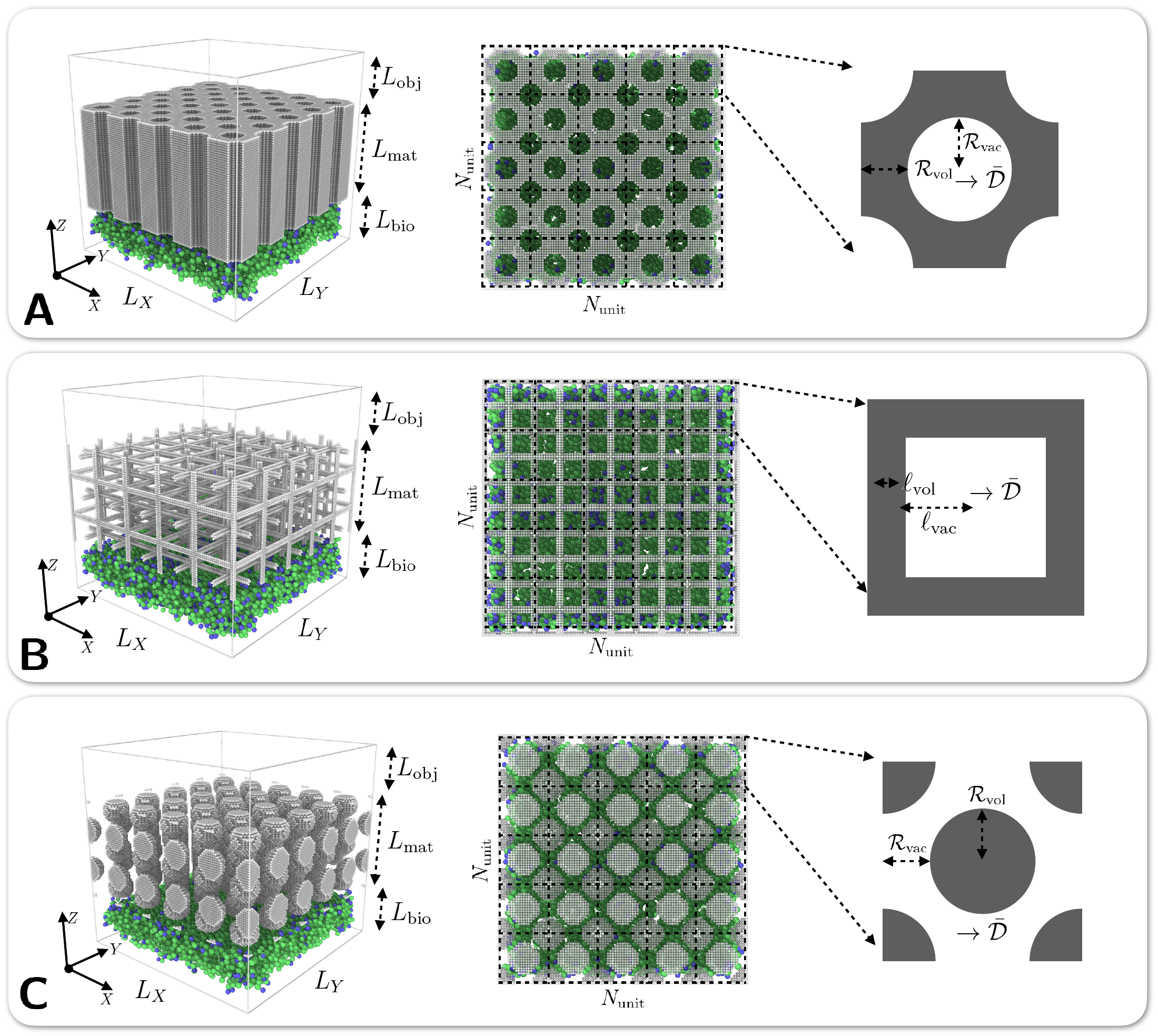}
    \caption{The schematic illustration for the three different porous materials formulations. The porous materials are treated as repeated elements of unit cells, and the number of unit cells per length is $N_{\rm unit}$ (marked in the middle sub-figures), which is defined as a design variable in the optimization. For every unit cell, the dimensionless structure parameter, $\bar{\mathcal{D}}$, is defined to quantify the vacuum-solid region spatial ratio in a defined unit cell illustrated in the right sub-figures. \textbf{(A)} Two-dimensional porous membranes for biofilm transport. Note that we use the term ``two-dimensional'' to denote that there are no repeating unit cells in the third dimension, i.e., the Z axis. The design variables hence do not perturb the geometries in the third dimension. Bacteria cells are grown within the ``micro-pipelines'' within the membranes to the top region. The dimensionless variable writes $\bar{\mathcal{D}} = \mathcal{R}_{\rm vac}/ (\mathcal{R}_{\rm vac} + \mathcal{R}_{\rm vol})$, is defined as the radii ratio between the vacuum region and the overall region (vacuum + volumetric solid). \textbf{(B)} Lattice porous metamaterials for biofilm transport. Bacteria cells are grown within the porous region within the lattice microstructures to reach the top. The unit cell dimensionless variable takes the form $\bar{\mathcal{D}} = \ell_{\rm vac}/ (\ell_{\rm vac} + \ell_{\rm vol})$, is defined as the length ratio between the vacuum region and the overall region. \textbf{(C)} Non-convex three-dimensional porous media for biofilm transport. Bacteria cells are grown within the porous region within the porous media to reach the top. The unit cell dimensionless variable takes the form $\bar{\mathcal{D}} = \mathcal{R}_{\rm vol}/ (\mathcal{R}_{\rm vac} + \mathcal{R}_{\rm vol})$, is defined as the radii ratio between the volumetric region and the overall region. }
    \label{fig2}
\end{figure}
Here, we define three different cases to simulate the process of biofilm growth constrained within porous materials, inspired by experimental setup, literature results, and natural phenomena. The general schematic representing the numerical experimental setup is illustrated in Figure \ref{fig2}. From Equation (\ref{opt_prob_equation}), $\mathcal{N}_{\rm bio}$ is the number of bacteria cells in the top quarter region and denoted as the {\em objective growth region}, i.e., $L_{\rm obj}\times L_X\times L_Y$. The porous microstructures are defined in the {\em materials region}, i.e., $L_{\rm mat}\times L_X\times L_Y$. The initial bacteria cells are distributed in the {\em initial biomass region}, i.e., $L_{\rm bio}\times L_X\times L_Y$. $N_{\rm unit}$ are formulated differently based on the ``dimension'' of the problem, where for the porous membrane (Figure \ref{fig2} \textbf{A}) $N_{\rm unit}$ is only defined in the X-Y plane. For lattice metamaterials and non-convex porous media, it is defined in all the X, Y, and Z directions. $\bar{\mathcal{D}}$ are defined within the unit cells. Here, $L_X = L_Y = 50\mu\rm m$, $L_{\rm bio} = L_{\rm obj} = 12.5\mu\rm m$, and $L_{\rm mat} = 25\mu\rm m$. The three cases are as follows:


\begin{itemize}
    \item {\bf Porous Membranes.} Biofilm growth and flow constrained in a microchannel are widely applied and studied by the microfluidics communities and their wide applications spanning from energy, biosensing, and many others \cite{biofilm_chanel_flow_review, channel_flow_anode_energy}. Many numerical \cite{channel_flow_numerical_study, channel_flow_numerical_study_2} and theoretical \cite{biofilm_chanel_theoretical_ref} studies also explored the mechanisms of biofilm growth and flow in microchannels. Here, our numerical implementations for channeled biofilm growth are mainly inspired by the simulation setup by Aspa et al. \cite{channel_flow_numerical_study_2}, where cylinder-shaped convex pores are ``drilled'' in the solid materials to create channels for biofilm to grow within (Figure \ref{fig2} \textbf{A}). The morphology of the unit cell is shown in the right subfigure in Figure \ref{fig2} \textbf{A}: the radius of the hole (vacuum area) is denoted as $\mathcal{R}_{\rm vac}$ and the length of the residual solid body (the volumetric part, equals to half length of the unit cell minus $\mathcal{R}_{\rm vac}$) is denoted as $\mathcal{R}_{\rm vol}$. The dimensionless variable can then be computed as $\bar{\mathcal{D}} = \frac{\mathcal{R}_{\rm vac}}{\mathcal{R}_{\rm vac} + \mathcal{R}_{\rm vol}}$. In this scenario, the range of the dimensionless variable is defined as $\bar{\mathcal{D}}\in [0.1,0.9]$ ($\bar{\mathcal{D}}_{\rm LB}$ and $\bar{\mathcal{D}}_{\rm UB}$ in Equation (\ref{opt_prob_equation})). Our optimized results from designing porous channels (or 2D porous membranes) could potentially be deployed for biofilm transport and utilization as ELM, as these topologies are easy to fabricate. We also benchmark the effects of the acquisition function in sampling the design space from BO (Section \ref{subsubsec_acqfunc}), in which we also characterize the design space from the sampling perspective that could guide general structural design optimizations.
    \item {\bf Lattice Metamaterials.} In recent years, there has been a huge growth in studies of the designs \cite{renee_dl_design, carlos_natcomm_design} and properties \cite{wendy_nano_architectured, carlos_pnas} of mechanical metamaterials (or synonymously architectured materials). However, their potential applications in biomass storage and transport are rarely explored, with very few works concerning their potential use as biofilm carriers \cite{lattice_mat_biofilm_ref_MSThesis, lattice_mat_biofilm_ref_energy} and related properties \cite{lattice_mat_biofilm_ref_biomed}. Here, we hope to use our simulations to fill in this gap and bring new insights into the possibilities of using lattice metamaterials for biofilm storage and transport. The unit cell of such metamaterials is shown in the right subfigure Figure \ref{fig2} \textbf{B}: the half length of the vacuum area is denoted as $\ell_{\rm vac}$, and the edge length of the solid volumetric part is denoted as $\ell_{\rm vol}$, where the dimensionless variable is defined as $\bar{\mathcal{D}} = \frac{\ell_{\rm vac}}{\ell_{\rm vac} + \ell_{\rm vol}}$. The range of the dimensionless variable is defined as $[0.1, 0.5]$.
    \item {\bf Non-convex Porous Media.} Inspired by the fact that biofilms were mostly found in natural habitats where they were constrained in pseudo- or spherical solid bodies \cite{natcomm_bacteria_hopping_porous, biofim_porous_3D_ref, pnas_biofilm_competetion, pnas_porous_flow}, we propose the simulation scenario where biofilm grows in nonconvex solid bodies shown in Figure \ref{fig2} \textbf{C}. The simulations were mainly inspired by the study of Dehkharghani et al. \cite{commphys_scale_pore_effect} and Bhattacharjee \& Datta \cite{natcomm_bacteria_hopping_porous}, where we use BO as a tool to sample the scale effect studied in \cite{commphys_scale_pore_effect} which defined a similar 3D porous packing of solid spherical bodies in \cite{natcomm_bacteria_hopping_porous}. The dimensionless variable is defined as the radii ratio between the solid spheres and the overall unit cell lengths (right subfigure in Figure \ref{fig2} \textbf{C}): $\bar{\mathcal{D}} = \frac{\mathcal{R}_{\rm vol}}{\mathcal{R}_{\rm vol} + \mathcal{R}_{\rm vac}}$. The range of the dimensionless variable is defined as $[0.5, 1.2]$. While such porous structures may not be easily fabricated, it closely resembles natural structures found in nature, hence we hope to optimize this structure with the BO sampling to investigate biofilm transport in such porous environments.
\end{itemize}
We use the porous membrane case to first evaluate the acquisition functions used and apply the BO for 500 iterations each. For the lattice metamaterials case, due to the high computation burden of the simulation, we only apply BO for 300 iterations with only the EI acquisition function. For the porous media case, we apply BO for 500 iterations with only the EI acquisition function. For all three cases, we conduct simulations to examine the accuracy of the GPR approximated design space at the maximal point in the visualized reconstructed design space, as well as at a randomly selected point in the design space to serve as a control.

\section{Results \& Discussion\label{sec_resdis}}

\subsection{Porous Membranes\label{subsec_2dmat}}

Three questions may naturally arise from the simulation-based Bayesian optimizations:\begin{enumerate}
    \item[(I)] Just observing the changes in the objectives may not be comprehensive enough to estimate whether both the acquisition functions are sampling toward the ``correct'' directions, i.e., whether the sampling directions are moving toward higher objective values which is the design goal.\item[(II)] Can we generally verify the accuracy of the design space approximated by GPR?\item[(III)] What are the exact geometries represented by the changing variables?
\end{enumerate}

\begin{figure}[htbp]
    \centering
    \includegraphics[width=42em]{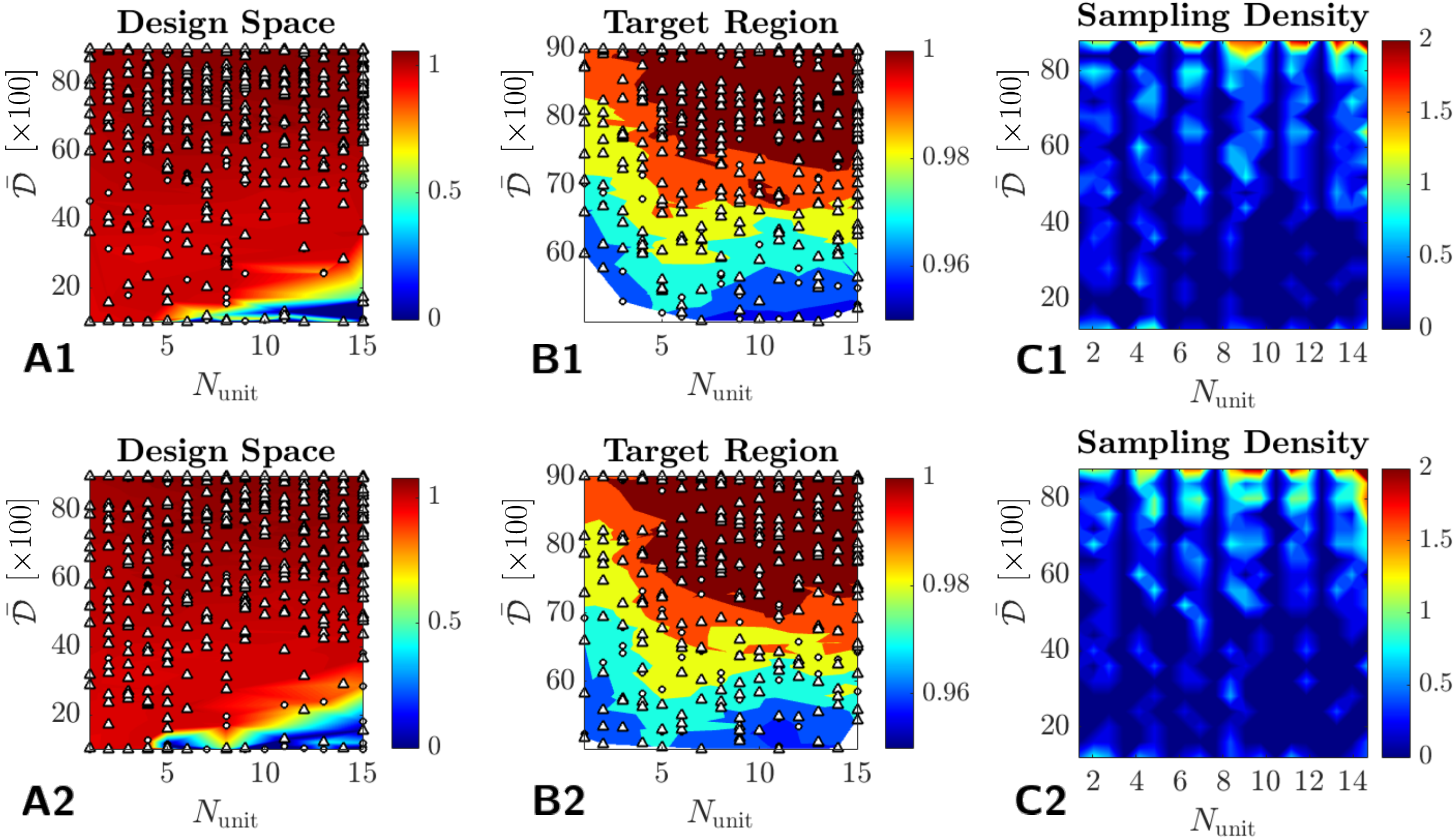}
    \caption{The design space reconstruction (visualized in normalized values) and sampling density maps by the two different acquisition functions for 2D porous membrane design case. Here, \textbf{(A1}$\sim$\textbf{C1)} stand for the design space surrogate and sampling density map from the EI acquisition function, and \textbf{(A2}$\sim$\textbf{C2)} stand for those by UCB acquisition function. Note that for subfigures \textbf{A}, the white dots are visualized in three batches: the first batch represents the first 300 iterations, visualized in small circular dots, the mid-100 iterations are visualized as squared-shaped dots, and the last 100 iterations are visualized in large triangular dots, which are the easiest to be identified. For subfigure \textbf{B}, the visualization of the first two batches remains the same, whereas the last batch set contains different evaluations and is marked still in triangular dots. For details please see the main text. The primary goal is to characterize the sampling density map through the morphology of the sampling dots in the reconstructed design space. \textbf{(A1)} The reconstructed design space by EI acquisition function. \textbf{(B1)} Zoomed view toward the target design region from subfigure \textbf{A1}, where $N_{\rm unit}\in [5,15]$, and $\bar{\mathcal{D}}\times100\in [50,100]$. \textbf{(C1)} The normalized sampling density map for the EI acquisition function, visualizing the density of the choice of the design variables in the optimization processes. \textbf{(A2)} The reconstructed design space by UCB acquisition function. \textbf{(B2)} Zoomed view toward the target design region from subfigure \textbf{A2}, where $N_{\rm unit}\in [5,15]$, and $\bar{\mathcal{D}}\times100\in [50,100]$. \textbf{(C2)} The normalized sampling density map for the UCB acquisition function, visualizing the design variables' densities in the optimization processes. }
    \label{fig4}
\end{figure}

\begin{figure}[htbp]
    \centering
    \includegraphics[width=47em]{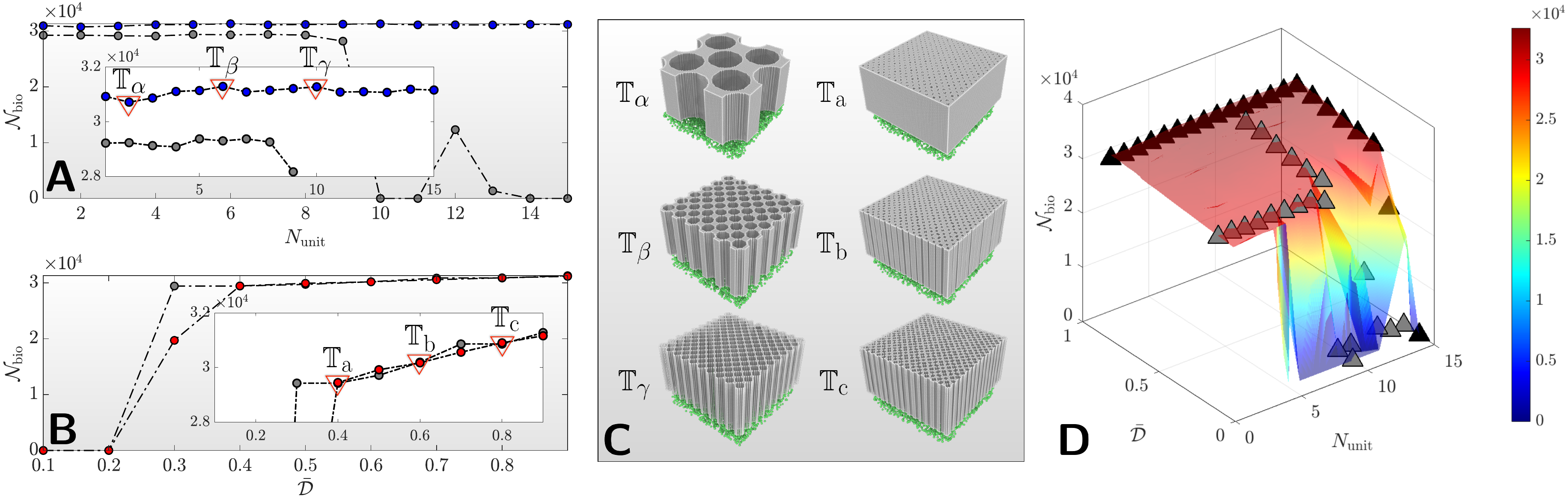}
    \caption{Design space characterization for the GPR reconstructed design space and topologies extraction from the characterization processes for the 2D porous membrane design case. \textbf{(A)} Characterization of the design variable $N_{\rm unit}$ with different fixed values of $\bar{\mathcal{D}}$. Note that the blue circular dots correspond to the black triangular dots, and the grey circular dots corresponds to the black triangular dots, in subfigure \textbf{D}. The blue and red circular dots are the characterization tests informed by qualitative observation of the GPR reconstructed design space to approximate the optimal design (i.e., maximal point), and the grey dots are random tests to benchmark our characterization informed by the observations. The zoomed view describes the detailed differences between the two sets of characterization simulations, in which three sets of membrane topologies are selected and highlighted in red triangular plots, nominated as $\mathbb{T}_\alpha$, $\mathbb{T}_\beta$, and $\mathbb{T}_\gamma$, respectively. \textbf{(B)} Design variable characterization for $\bar{\mathcal{D}}$ compared with random benchmark test marked in red and grey dots, respectively. The zoomed view describes the detailed differences between the two sets of characterization simulations, in which three sets of membrane topologies are selected and highlighted in red triangular plots, nominated as $\mathbb{T}_{\rm a}$, $\mathbb{T}_{\rm b}$, and $\mathbb{T}_{\rm c}$, respectively. \textbf{(C)} Extracted porous membranes' topologies ($\mathbb{T}_\alpha\sim\mathbb{T}_\gamma$ \& $\mathbb{T}_{\rm a}\sim\mathbb{T}_{\rm c}$) from characterizing both the design variables $N_{\rm unit}$ and $\bar{\mathcal{D}}$ corresponding to the selections in subfigures \textbf{A} \& \textbf{B}. \textbf{(D)} The characterization data match with the GPR reconstructed design spaces from both the EI and UCB acquisition function. The black triangular dots are the characterization informed by observation from the GPR reconstructed design space towards the maximal value. The grey triangular dots are randomly selected test points to benchmark the observation-informed characterizations.}
    \label{fig5}
\end{figure}

Note that these three questions are fundamental in our following analyses for different materials design cases. Here, to answer Question (I), we visualize the sampling process during the optimizations and characterize them with the overall sampling density (Figure \ref{fig4}). To answer Questions (II) \& (III), we characterize the approximated design space using simulations and visually show the general trends captured by the approximated models and simulation points (Figure \ref{fig5}). We then further visualize the geometries extracted from the characterization simulations.

Figure \ref{fig4} \textbf{A1} \& \textbf{A2} visualize the overall reconstructed design spaces updated by EI and UCB acquisition functions. Note that the dimensionless variable $\bar{\mathcal{D}}$ is multiplied by 100 in the visualizations for ease of analysis. The two different acquisition functions all approximated the same trend: there is a large objective gradient changing from the bottom-right corner. Physically, this would indicate that when the pores' radii ($\mathcal{R}_{\rm vac}$ in Figure \ref{fig2} \textbf{A}) are small and the unit cell numbers ($N_{\rm unit}$) are generally larger, the biofilm transport capabilities of the porous structures decrease. Also, the objective values are qualitatively higher with higher $\bar{\mathcal{D}}$ values, i.e., $\bar{\mathcal{D}}\gtrapprox 0.5$. We hence visualize the ``upper design space'' in Figure \ref{fig4} \textbf{B1} \& \textbf{B2}, in which the region $\bar{\mathcal{D}}\in [0.5,0.9]$ are visualized. The objective values are higher in the ``top-right'' corner of the design space, where both the sampling points' density and normalized objective values are higher. By directly visualizing the (normalized) sampling density (Figure \ref{fig4} \textbf{C}), we observe that the sampling density distribution basically overlaps with our observations on the design space: there are higher sampling densities toward the top-right corners (i.e., higher $\bar{\mathcal{D}}$ and $N_{\rm unit}$ values) characterized by both acquisition functions. Combining both Figure \ref{fig4} \textbf{A}, \textbf{B}, \& \textbf{C}, we deduce that both the reconstructed design spaces and the sampling densities tell us for the case of the porous membrane, the $\bar{\mathcal{D}}$ \& $N_{\rm unit}$ are positively correlated to the biofilm transportability $\mathcal{N}_{\rm bio}$ towards the target region. Here, the EI acquisition function samples 407 points in the ``upper design space'' (Figure \ref{fig4} \textbf{B1}), and the UCB acquisition function samples 373 points (Figure \ref{fig4} \textbf{B2}). If we define design space in Figure \ref{fig4} \textbf{B} as the target region, the EI acquisition sampling technique is 9.12\% more efficient relative to the acquisition function. If we only look at the last 100 iterations from the BO, the EI acquisition function samples 87 points in the target region, and the UCB acquisition function samples 85 points. Compared with a uniformly distributed grid search method, the EI acquisition function is 74\% more efficient and the UCB acquisition function is 70\% more efficient. The EI acquisition function is 2.35\% more efficient than the UCB acquisition by estimating the last 100 design space samples in the target region. To cross-verify these cross-validated observations from a more quantitative perspective and answer our Questions (II) \& (III), we characterize the design space using additional simulations (Figure \ref{fig5}).

Figure \ref{fig5} \textbf{A} \& \textbf{B} show the general and zoomed views of the design space characterizations. We compare the selected characterization simulations (in colored dots) and randomly selected simulations (in grey dots) to verify the effect of the design variables ($\bar{\mathcal{D}}$ \& $N_{\rm unit}$) to the target bacteria cell numbers $\mathcal{N}_{\rm bio}$. Here, the blue dots and grey dots in Figure \ref{fig5} \textbf{A} are extracted based on $\bar{\mathcal{D}} = 0.9$ and 0.2, respectively. The blue dots and grey dots in Figure \ref{fig5} \textbf{B} are extracted based on $N_{\rm unit} = 15$ and 10, respectively. The $\bar{\mathcal{D}}$ and $N_{\rm unit}$ values for blue and red dots are selected based on observations from Figure \ref{fig4} as our guess for the porous materials' geometries that contain the highest objective value. The $\bar{\mathcal{D}}$ and $N_{\rm unit}$ values for the grey dots are randomly selected to compare with our observational guess. We then directly visualize the points from the characterization simulations on the GPR reconstructed design space in Figure \ref{fig5} \textbf{D}. It can be observed that the characterization simulation tests fit well with the GPR-approximated design space as both the black and grey dots overlap well with the surface contours. We then pick a series of representative points from the characterization simulations and directly visualize them in Figure \ref{fig5} \textbf{C} (the points marked in red triangles in Figure \ref{fig5} \textbf{A} \& \textbf{B}) denoted as $\mathbb{T}_\alpha\sim\mathbb{T}_\gamma$ and $\mathbb{T}_{\rm a}\sim\mathbb{T}_{\rm c}$. In Figure \ref{fig5} \textbf{A}, $\mathbb{T}_\alpha$ is evidently smaller than that of $\mathbb{T}_\beta$ and $\mathbb{T}_\gamma$, and we can further deduce that the porous membrane with larger pores does not necessarily enhance the transportability of the porous materials, which is not intuitive. We propose that the repulsive mechanical forces, due to the wall of the pores, drive the new bacteria cells to grow towards the upper region. When the radii of the pores are too large, such reactive forces acting on the bacteria cells are not as strong due to fewer contacts with the cells. Moreover, it can be observed from Figure \ref{fig5} \textbf{B} that for $N_{\rm unit}=10$ \& 15, the effects of the dimensionless variable $\bar{\mathcal{D}}$ on the objective $\mathcal{N}_{\rm bio}$ are similar, where there are sudden increases of the objective between $\bar{\mathcal{D}} \in[0.2, 0.4]$. 

Based on our analyses in the case of porous membranes, we further deduce that the EI acquisition function outperforms the UCB acquisition function by estimating the objective variance, the mean objective values, and sampling improvements over the design space. We also observe that with larger relative radii of the pores and more unit cells per side, the transportability of porous materials to biofilms is then higher, from analyzing the design space. Therefore, subsequently, we will adopt only the EI acquisition function and conduct further analyses for lattice and 3D nonconvex porous media (Section \ref{subsec_3dmat}).



\subsection{Lattice and Porous Materials\label{subsec_3dmat}}

\begin{figure}[htbp]
    \centering
    \includegraphics[width=45em]{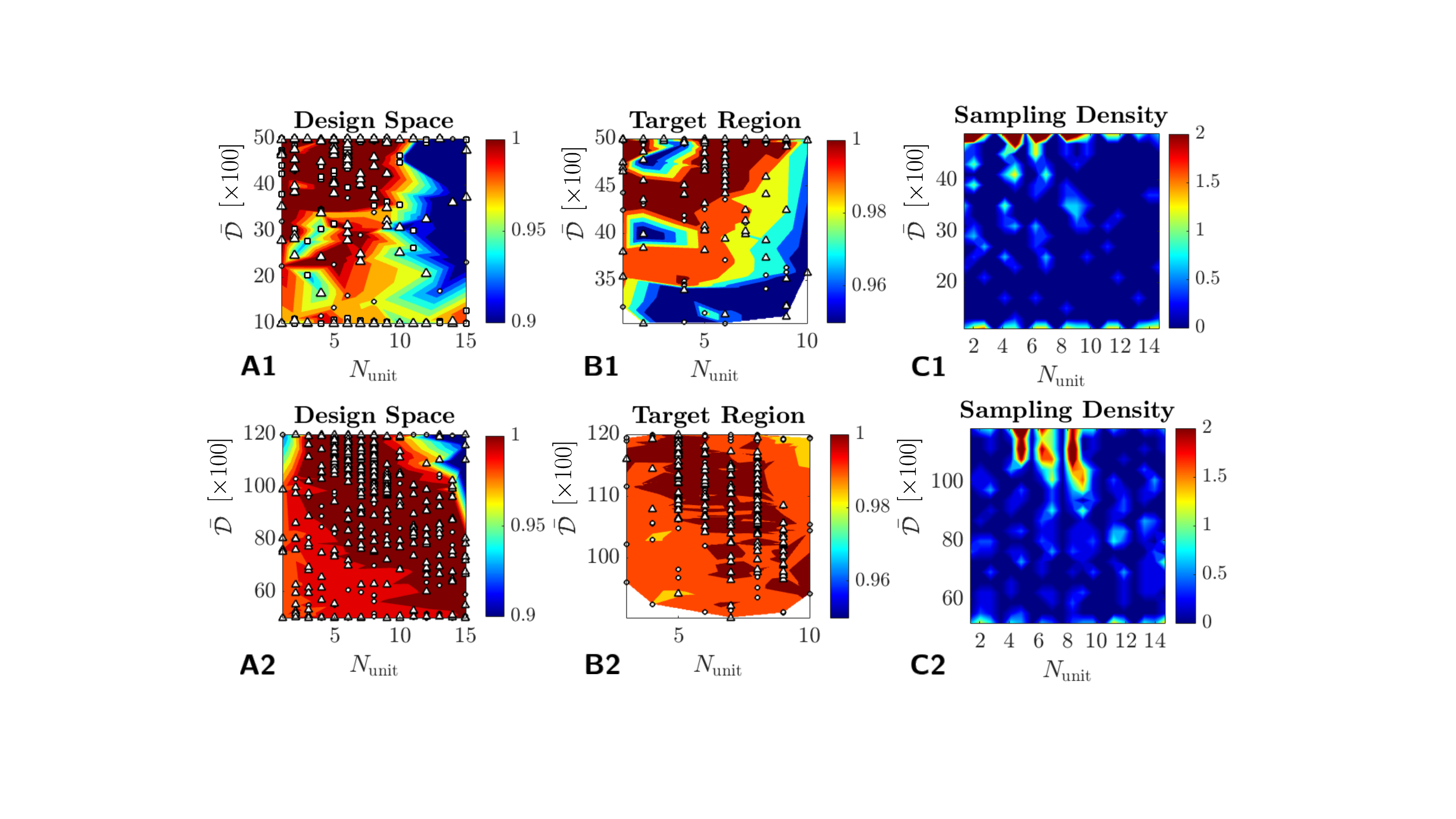}
    \caption{The design space reconstruction (visualized in normalized values) and sampling density maps by the two different acquisition functions for lattice metamaterials \textbf{(A1}$\sim$\textbf{C1)} and 3D porous media \textbf{(A2}$\sim$\textbf{C2)}, updated by the EI acquisition function. The morphologies of the white dots are separated into three different batches. \textbf{(A1)} The reconstructed design space by EI acquisition function. The first batch represents the first 100 iterations, visualized in small circular dots, the mid-100 iterations are visualized as squared-shaped dots, and the last 100 iterations are visualized in large triangular dots. \textbf{(B1)} Zoomed view toward the target design region from subfigure \textbf{A1}, where $N_{\rm unit}\in [1,10]$, and $\bar{\mathcal{D}}\times100\in [30,50]$. The first batch represents the first 100 iterations, visualized in small circular dots, the mid-50 iterations are visualized as squared-shaped dots, and the large triangular dots represent the rest visualizations. For details please see the main text. \textbf{(C1)} The normalized sampling density map for the EI acquisition function for the lattice metamaterials design case, visualizing the density of the choice of the design variables in the optimization processes. \textbf{(A2)} The reconstructed design space by EI acquisition function. The first batch represents the first 300 iterations, visualized in small circular dots, the mid-100 iterations are visualized as squared-shaped dots, and the last 100 iterations are visualized in large triangular dots. \textbf{(B2)} Zoomed view toward the target design region from subfigure \textbf{A2}, where $N_{\rm unit}\in [1,10]$, and $\bar{\mathcal{D}}\times100\in [90,120]$. The first batch represents the first 300 iterations, visualized in small circular dots, the mid-50 iterations are visualized as squared-shaped dots, and the large triangular dots represent the rest. \textbf{(C2)} The normalized sampling density map for the EI acquisition function for the 3D porous media design case, visualizing the design variables' densities in the optimization processes. }
    \label{fig6}
\end{figure}
\begin{figure}[htbp]
    \centering
    \includegraphics[width=47em]{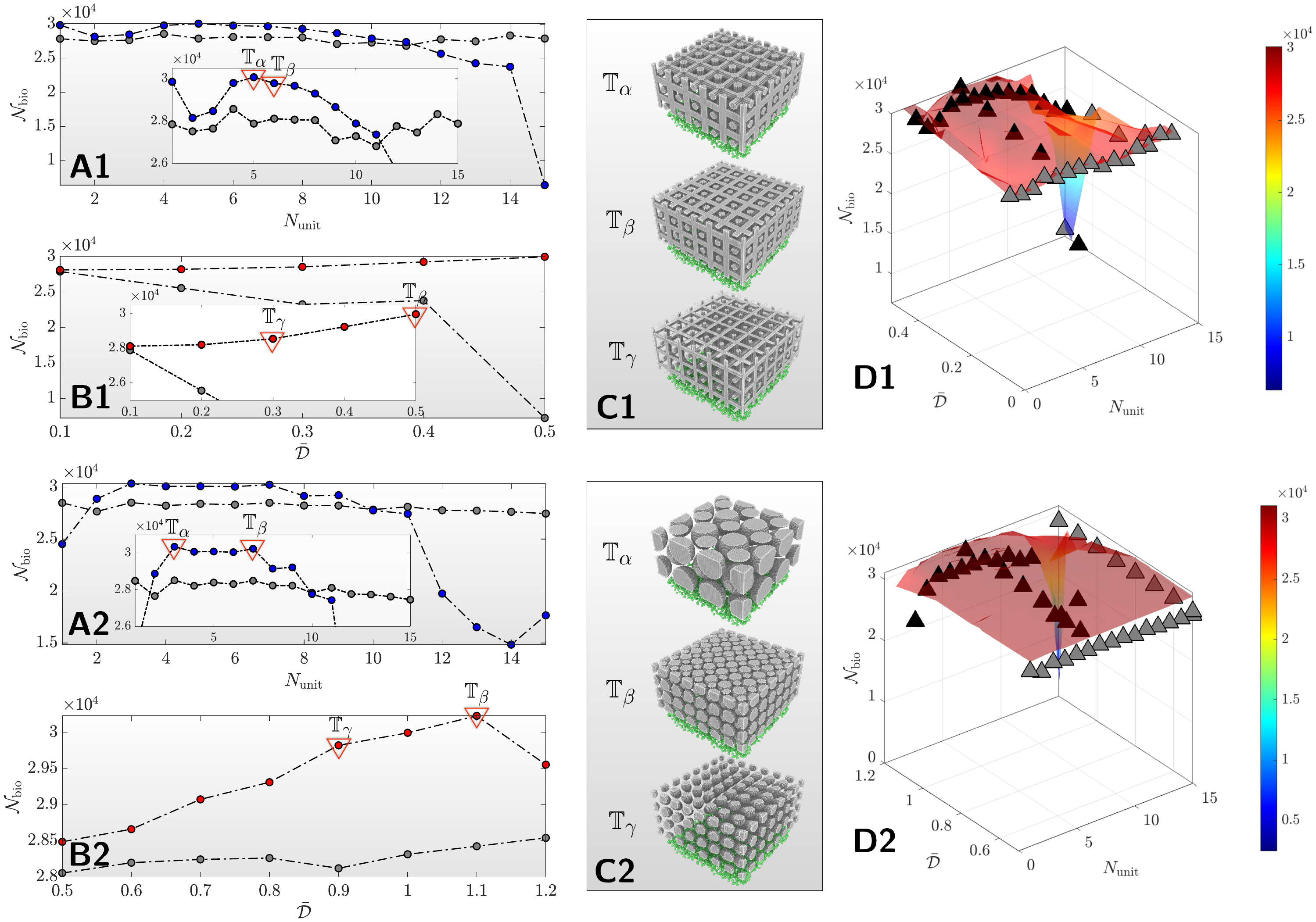}
    \caption{Design space characterization for the Gaussian process regression (GPR) reconstructed design space and topologies extraction from the characterization processes for both the lattice metamaterials and 3D porous media design optimization. \textbf{(A1)} Characterization of the design variable $N_{\rm unit}$ with different fixed values of $\bar{\mathcal{D}}$. Note that the blue circular dots correspond to the black triangular dots, and the grey circular dots corresponds to the black triangular dots, in subfigure \textbf{D1}. The blue and red circular dots are the characterization tests informed by qualitative observation of the GPR reconstructed design space to approximate the optimal design (i.e., maximal point), and the grey dots are random tests to benchmark our characterization informed by the observations. The zoomed view describes the detailed differences between the two sets of characterization simulations, in which three sets of membrane topologies are selected and highlighted in red triangular plots, nominated as $\mathbb{T}_\alpha$, and $\mathbb{T}_\beta$, respectively. \textbf{(B1)} Design variable characterization for $\bar{\mathcal{D}}$ compared with random benchmark test marked in red and grey dots, respectively. The zoomed view describes the detailed differences between the two sets of characterization simulations, in which three sets of membrane topologies are selected and highlighted in red triangular plots, nominated as $\mathbb{T}_\beta$, and $\mathbb{T}_\gamma$, respectively ($\mathbb{T}_\beta$ is the same topology as in subfigure \textbf{A1}). \textbf{(C1)} Extracted porous membranes' topologies ($\mathbb{T}_\alpha\sim\mathbb{T}_\gamma$) from characterizing both the design variables $N_{\rm unit}$ and $\bar{\mathcal{D}}$ corresponding to the selections in subfigures \textbf{A1} \& \textbf{B1}. \textbf{(D1)} The characterization data match with the GPR reconstructed design spaces from the EI acquisition function. The black triangular dots are the characterization informed by observation from the GPR reconstructed design space towards the maximal value. The grey triangular dots are randomly selected test points to benchmark the observation-informed characterizations. \textbf{(A2)} Characterization of the design variable $N_{\rm unit}$ with different fixed values of $\bar{\mathcal{D}}$. Visualization details are the same as in subfigure \textbf{A1}. \textbf{(B2)} Design variable characterization for $\bar{\mathcal{D}}$ compared with random benchmark test marked in red and grey dots, respectively. Visualization details are the same as in subfigure \textbf{B1}, except there is no zoomed view since the range for the objective $\mathcal{N}_{\rm bio}$ are already within a small range. \textbf{(C2)} Extracted porous membranes' topologies ($\mathbb{T}_\alpha\sim\mathbb{T}_\gamma$) from characterizing both the design variables $N_{\rm unit}$ and $\bar{\mathcal{D}}$ corresponding to the selections in subfigures \textbf{A2} \& \textbf{B2}. \textbf{(D2)} The characterization data match with the GPR reconstructed design spaces from the EI acquisition function. Visualization details are the same as in subfigure \textbf{D1}.}
    \label{fig7}
\end{figure}

Figure \ref{fig6} shows the reconstructed design space and the sampling process along with the sampling density updated by the EI acquisition function. It can be observed from Figure \ref{fig6} \textbf{A1} that the reconstructed design space from 300 evaluations is much more nonconvex compared with that of the 2D porous membrane (Figure \ref{fig4} \textbf{A}) and porous media (Figure \ref{fig6} \textbf{A2}), but the sampling is more concentrated toward the mid-top region ($N_{\rm unit}\approx 0.5$ \& $\bar{\mathcal{D}}\in[0.4,0.5]$). Figure \ref{fig6} \textbf{B1} visualizes this subregion ($N_{\rm unit}\in [1,10]$ \& $\bar{\mathcal{D}}\in [0.3,0.5]$), in which by qualitative estimation one deduces that there are more sampling points around $N_{\rm unit}=6$ and $\bar{\mathcal{D}} = 0.5$. Comparing the reconstructed design space and the sampling density (Figure \ref{fig6} \textbf{C1}), one observes that the general trends of the sampling density and the reconstructed design space overlap well, where we thence pick $N_{unit}=6$ and $\bar{\mathcal{D}}=0.45$ for further simulations based on qualitative observations (Figure \ref{fig7} \textbf{1}). Figure \ref{fig6} \textbf{A2} shows that the reconstructed design space is shaped like a ``tilted wave'' --- the higher objective values are distributed along the ``cross-split'' across the design space coordinates. By observing Figures \ref{fig6} \textbf{A2} \& \textbf{C2}, we deduce that the sampling density is more biased towards the ``upper design space''. Hence, we only extract the zoomed view of the top-mid design space in Figure \ref{fig6} \textbf{B2} ($N_{\rm unit}\in [1,10]$ \& $\bar{\mathcal{D}}\in[0.9,1.2]$). From Figure \ref{fig6} \textbf{B2}, we pick $N_{\rm unit} = 7$ and $\bar{\mathcal{D}} = 1.1$ to conduct characterization tests in Figure \ref{fig7} \textbf{2}.

To estimate the effect of the acquisition function on the sampling of the design space, we also estimate the spatial distribution of the last 100 iterations within the target design space (or target region), where the target regions are defined based on the zoomed design space in Figure \ref{fig6} \textbf{B} ($N_{\rm unit}\in [1,10]\ \&\ \bar{\mathcal{D}}\in [0.3,0.5]$ for lattice metamaterials in Figure \ref{fig6} \textbf{B1}; $N_{\rm unit}\in [1,10]\ \&\ \bar{\mathcal{D}}\in [0.9,1.2]$ for 3D porous media in Figure \ref{fig6} \textbf{B2}). For the lattice metamaterials, there are 62 points sampled in the target region, which is 92.89\% higher than the uniform distribution of 100 points with assumed grid search methods (32.14 points in the target region). For the 3D porous media, there are 89 points sampled in the target region, which is 223.04\% more efficient than the uniformed sampled 100 points (27.55 points in the target region). In summary, BO exhibits an outstanding ability for sampling towards the target design goal for both porous structure cases.

Other than our selected characterization tests, we also randomly pick two additional characterization tests as controls for characterizing $N_{\rm unit}$ and $\bar{\mathcal{D}}$ for each porous materials design case, respectively. For designing the lattice metamaterials, we pick $\bar{\mathcal{D}} = 0.1$ (Figure \ref{fig7} \textbf{A1}) and $N_{\rm unit}=15$ (Figure \ref{fig7} \textbf{B1}), and for 3D porous media design, we pick $\bar{\mathcal{D}} = 0.5$ (Figure \ref{fig7} \textbf{A2}) and $N_{\rm unit}=15$ (Figure \ref{fig7} \textbf{B2}). It can be observed from Figure \ref{fig7} \textbf{A} \& \textbf{B} that the selected characterization tests generally capture the geometries of the highest objectives, where the blue and red dots exhibit higher values than the grey dots. Interestingly, for both porous materials cases, the topology corresponds to the highest objective value selected from the characterization tests for $\bar{\mathcal{D}}$ (Figure \ref{fig7} \textbf{B}), $\mathbb{T}_\beta$ are not the topology that contends the highest objective value by characterizing $N_{\rm unit}$ (Figure \ref{fig7} \textbf{A}). This indicates that our observational guess toward the highest objective is not fully accurate, where our characterization tests correct our initial guess and contends the porous structural topologies $\mathbb{T}_\alpha$. By observing Figure \ref{fig7} \textbf{D1} \& \textbf{D2} we observe that the characterization tests generally match well with the GPR approximated design space, indicating the effectiveness of the general data-driven design scheme. Notwithstanding, by comparing Figure \ref{fig7} \textbf{D1} and \textbf{D2} it is observed that the characterization tests match better with the GPR approximated design space for the lattice structures than the nonconvex porous materials. Both Figure \ref{fig7} \textbf{A}, \textbf{B}, \& \textbf{D} indicate the importance of additional qualitative characterizations but also prove the general accuracy of the GPR approximation. 

Finally, we extract the optimal design for each case as references. Table \ref{tab_para} shows the objective values ($\mathcal{N}_{\rm bio}$), their corresponding design variables ($N_{\rm unit}$ \& $\bar{\mathcal{D}}$), and the transformed characteristic length $\mathfrak{L}$ (in the unit of $\mu\rm m$) for all three cases benchmarked by a nonconfined pure biofilm growth in vacuum space. Very interestingly and unexpectedly, it is observed that all the optimal designs extracted from porous materials confined biofilm growth exhibit more bacteria cells in the target growth region than nonconfined biofilm growth in a vacuum space. The optimal designs of the 2D porous membrane, lattice metamaterials, and 3D porous media have 16\%, 7\%, and 11\%, more biofilms in the target growth region than the pure growth in the vacuum space, respectively. This confinement-induced biofilm growth may help us (1) better utilize biofilms as ELM and address the three points presented in the second paragraph in Section \ref{sec_intro}, and (2) potentially explain the natural phenomena described in the first paragraph in Section \ref{sec_intro}. We focus on this point to conduct a further comparison study in the following Section \ref{subsec_biomechanics}.

\begin{table}[htbp]
    \centering
    \begin{tabular}{c| c c c c}
        & $\mathcal{N}_{\rm bio}$ & $N_{\rm unit}$ & $\bar{\mathcal{D}}$ & $\mathfrak{L}$ $[\mu\rm m]$\\\hline
        2D porous membrane & 32655 & 10 & 0.1 & 0.5\\
        & 32655 & 11 & 0.1 & 0.45\\
        Lattice metamaterials & 30096 & 1 & 0.5 & 25\\
        3D porous media & 31152 & 7 & 1.1 & 0.71\\
        Unconfined space & 28086 & N/A & N/A & N/A\\\hline
    \end{tabular}
    \caption{The highest objective values and their corresponding design variables for different porous materials design cases, with transformed characteristic lengths in the unit of $\mu\rm m$. For 2D porous membrane, the characteristic length is defined as $\mathfrak{L}\equiv\mathcal{R}_{\rm vac}$. For lattice metamaterials, the characteristic length is defined as $\mathfrak{L}\equiv\ell_{\rm vol}$. For 3D porous media, the characteristic length is defined as $\mathfrak{L}\equiv\mathcal{R}_{\rm vol}$. ``Unconfined space'' denotes the case where there are no porous substrates on top of the initial bacteria cells and hence the cells grow in an unconfined space.}
    \label{tab_para}
\end{table}

\subsection{Biomechanics of Porous Transport\label{subsec_biomechanics}}

\begin{figure}[htbp]
    \centering
    \includegraphics[width=45em]{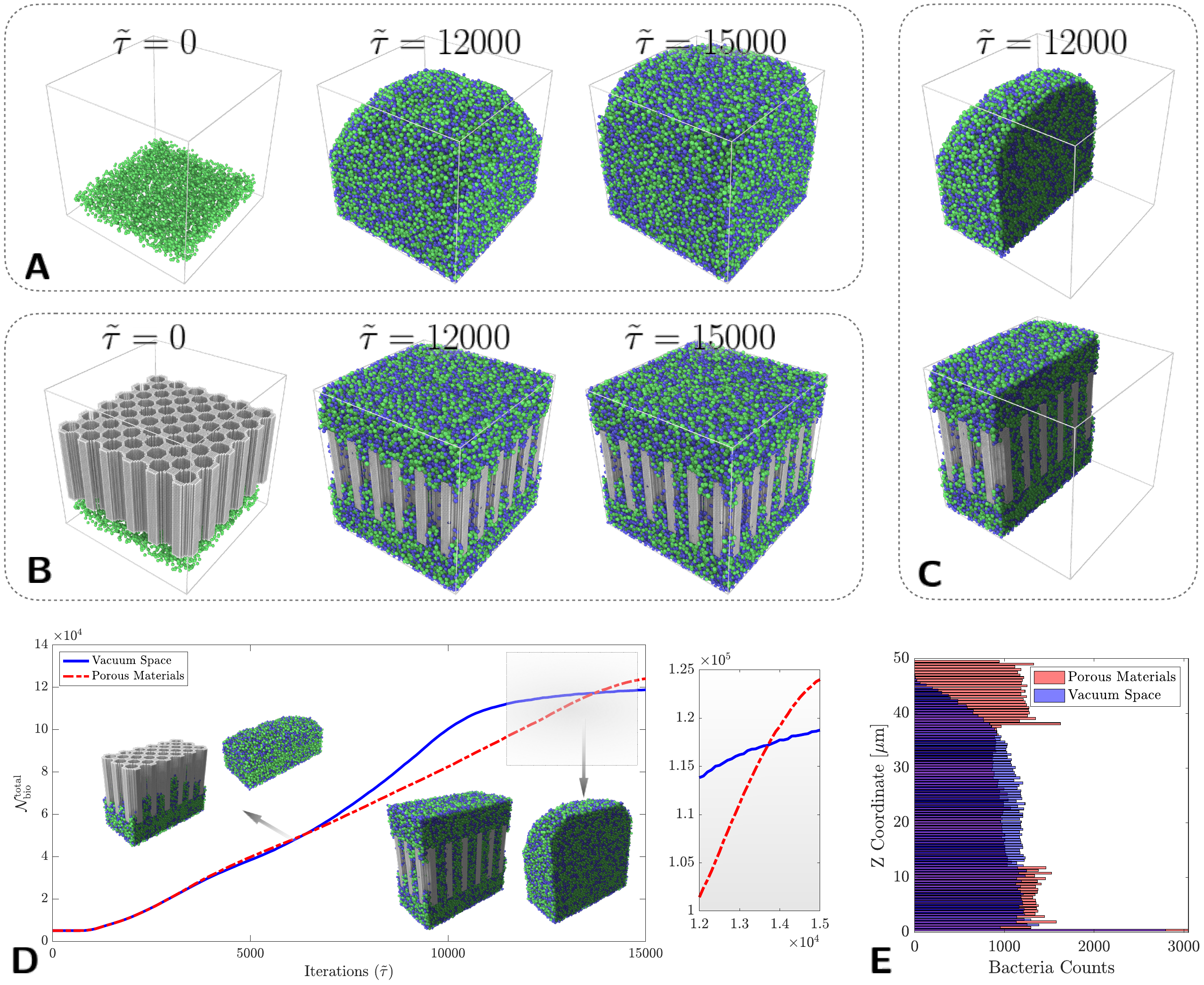}
    \caption{Comparison study for a single 2D porous membrane with vacuum biofilm growth case to unravel the biomechanics of porous materials induced biofilm growth. \textbf{(A)} The snapshots of the simulation of biofilm growth in pure vacuum space, where $\Tilde{\tau}$ is the simulation iteration step or can be treated as the pseudo-time. \textbf{(B)} The snapshots of the simulation of biofilm growth in the 2D porous membrane. \textbf{(C)} Slice view of snapshot $\Tilde{\tau} = 12000$ for both the 2D membrane and vacuum growth cases. \textbf{(D)} The accumulated bacteria cell numbers $\mathcal{N}_{\rm bio}$ along the iteration process, where the simulation snapshot of $\Tilde{\tau}=6000$ is indicated in the left top subfigure and $\Tilde{\tau}=13500$ is indicated in the bottom right subfigure. The solid blue line indicates the biofilm growth in vacuum space (without any porous materials) and the red dashed line indicates the biofilm growth in the 2D porous membrane for benchmarking. The zoomed view for $\Tilde{\tau}\in[12000,15000]$ is indicated in the right subfigure with a gradient-shaded background. \textbf{(E)} The bacteria cells' spatial distribution along the perpendicular direction (Z axis) at $\Tilde{\tau}=12000$, where the cell numbers are counted based on 100 interval slices visualized in bar plots. The blue bars indicate the vacuum space bacteria counts and the red bars indicate the bacteria counts in the 2D porous membrane. For details see the text. }
    \label{fig8}
\end{figure}

We perform further IbM simulations to potentially uncover the biomechanical mechanisms governing the optimal design. Figure \ref{fig8} shows the benchmark study of the biofilm growth in porous membranes and in unconfined space. We pick the case of a 2D porous membrane with $N_{\rm unit} = 6$ and $\bar{\mathcal{D}}$ for comparison with biofilm growth in unconfined space. Figure \ref{fig8} \textbf{A} \& \textbf{B} visualize the snapshots of the biofilm growth simulations, where $\Tilde{\tau}$ stand for the iteration number (or time steps), which can be converted to actual time as $t = 10\times\Tilde{\tau}\ \rm [s]$. Figure \ref{fig8} \textbf{C} visualizes the sliced view of the biofilm growth at $\Tilde{\tau}=12000$, to further explain confinement-induced biofilm growth. Figure \ref{fig8} \textbf{D} shows the change of the total bacteria cells $\mathcal{N}_{\rm bio}^{\rm total}$ along with the iterations $\Tilde{\tau}$, where the blue solid line denotes biofilm growth in unconfined space and the red dashed line denotes biofilm growth in the porous membrane. We observe two key moments that distinguish the overall biofilm growth: the first moment is at $\Tilde{\tau}\approx6000$ when the biofilm in the unconfined space (blue solid line) exceeds that of in the porous materials (red dashed line), and the second moment is at $\Tilde{\tau}\approx13500$ when the biofilm in the porous materials (red dashed line) exceeds that of in the vacuum space (blue solid line). The sliced views of the two moments ($\Tilde{\tau}=6000$ \& $\Tilde{\tau}=13500$) are visualized and indicated by shaded arrows. To quantitatively understand the mechanisms of confinement-induced biofilm growth and transport, we compute the biofilm cell numbers distribution along the Z-axis by counting through 100 slices at $\Tilde{\tau}=12000$ (detailed analysis can be found in ESI of Ref. \cite{zhai_acs_biofilm}) and visualize the results in Figure \ref{fig8} \textbf{E}, in direct correspondence with the illustration in Figure \ref{fig8} \textbf{C}. The blue bars indicate the accumulative bacteria counts for biofilm growth in vacuum space and the red bars indicate that of the porous materials.

It can be observed from Figure \ref{fig8} \textbf{A} \& \textbf{B} that the biofilm is more densely compacted in the target growth region through the porous materials compared with the growth in the unconfined space. From the sliced view in Figure \ref{fig8} \textbf{C}, we may hence propose a qualitative explanation for our observation: the existence of the porous material takes a certain amount of volume, which pushes the biofilm to grow upwards to occupy more space. While intuitive that this should occur, it is interesting to note that the growth proceeds in a nonlinear fashion. To break down this process in more detail, Figure \ref{fig8} \textbf{D} shows that after $\Tilde{\tau}\approx 6000$ the existence of the porous materials first suppress the biofilm growth, as $\mathcal{N}_{\rm bio}^{\rm total}$ for unconfined space (solid blue line) first increases nonlinearly with larger values than that of porous materials (dashed red line). But after the biofilm has grown extensively in the target growth region ($\Tilde{\tau}\approx13500$), the pores in the porous materials can be treated as ``channels'' that enhance the growth and transport of biofilms. This finding is significant in the sense that the effects of porous materials on the overall growth of biofilms change in different stages of the growth processes within the pores. Based on these comprehensive qualitative analyses, Figure \ref{fig8} \textbf{E} offers quantitative evidence that the porous structure helps to speed up biofilms' upward growth by taking up volumetric spaces --- the biofilm accumulation within the porous materials spatial range ($Z\in[12.5,37.5]\mu\rm m$) for porous materials (red bars) are evidently smaller than that of vacuum space (blue bars). Based on the bacteria cell numbers count from 100 slices, the bacteria cell counts within the porous region for the porous membrane and that of the unconfined space are 48643 and 58482, respectively. The unconfined space contains 20\% more bacteria cells than when constrained by the porous membrane. The target growth region bacteria counts for porous membrane and unconfined space are 31404 and 13764, respectively, where the porous membrane contains 128\% more cells than the cells grown in the unconfined space. The data not only verifies our qualitative explanations that the porous membrane facilitates biofilm growth by taking up volumetric space but also further explains how the porous membrane increases the overall cell counts --- the pores behave like channels that transport biofilms to the target region so that the bacteria count in the target growth region for porous membrane are significantly larger than that in unconfined spaces.

{\color{black} Based on our previous discussions and our inspiration in Section \ref{sec_intro}, an interesting question arose: how can we connect observations from our numerical experiments to natural phenomena? The ``porosity-induced growth'' can be related to both the biophysical behavior of biofilms and engineering control strategies. From the perspective of biofilm's biophysical behavior, our numerical experiments unveil a novel understanding that biofilm growth within constrained environments is shaped by the responsive forces from the porous scaffolds that are likely to stimulate the biofilms' growth, as the overall bacteria cells increase compared with the vacuum space benchmark (Figure \ref{fig8}). Moreover, the transition of porosity-induced biofilm growth happened at a specific time (Figure \ref{fig8} \textbf{D}), suggesting that such a phenomenon may occur at a critical temporal moment in nature. From the engineering control perspective, the simulations demonstrated that vacuum volumetric spaces with certain channel-like patterns could guide biofilm growth in distinct directions. This observation can inspire further strategies to manufacture engineering structures for biofilm control. For instance, channels may be carefully crafted to control the pathways for biofilm growth to target biomass transport. The transition observed in Figure \ref{fig8} \textbf{D} suggests that one should consider the temporal effects when manufacturing engineering scaffolds to control and utilize biofilms.}


{\color{black}

\subsection{Limitations of the Framework\label{subsec_limitations}}

Despite the successful implementation and physical insights, there are still a few limitations of this computation-based framework. From the modeling perspective: First, we assume constant growth of biofilm, which may not universally describe the dynamical behavior of all the biofilms. Second, we assume a linear nutrient distribution to stimulate biofilm growth, which may not perfectly align with real-world scenarios --- biofilms may grow in nonlinear nutrient distributions in complex environments. Eventually, we assume the metamaterials are rigid and cannot account for nonlinear mechanical responses (e.g., viscoelasticity, hyperelasticity, and plasticity).

From the optimization and machine learning perspective: there are a few future possible improvements based on our limitations. First, the IbM simulations are still computationally expensive, and we hope to develop reduced-order models and/or data-driven surrogates for IbM models for fast function evaluations in BO. Second, there are currently no uncertainty bounds in our IbM computational models, and we are developing Monte Carlo-based uncertainty sampling to estimate the underlying parametric density distribution for efficiently formulating the optimization problem.

}

\section{Conclusions \& Outlook \label{sec_conclusion}}


We presented different designs of porous structures for enhanced biofilm transport and control using computational simulations and Bayesian optimization. We characterized the design optimization process, comprehensively analyzed the approximated design space, and provided in-depth physical insights from the optimization. We formulated three different types of porous structural materials for design optimization aiming to maximize the biofilms in the target growth region. For three different types of porous materials, the trends of the reconstructed design space matched well with the sampling density. For the 2D porous membrane, the variance of the overall samples by the UCB acquisition function was 32.08\% higher than that of the EI acquisition function; the mean objective of the overall samples by the EI acquisition function was 1.49\% higher than that of the UCB acquisition function. Given the predefined target region of higher sampled densities, the EI acquisition function was 2.35\% more efficient than the UCB acquisition function compared with uniformly distributed grid search methods by estimating the last 100 sampling points. The GPR approximated design spaces matched well with the selected characterization tests. Using only the EI acquisition function, we conducted the design space characterization for lattice metamaterials and porous media under the same procedure. For the lattice metamaterials, by observing the last 100 samples in the predefined target design space, BO was 92.89\% more efficient than the uniform grid search. For the 3D porous media, there were 223.04\% more sampled points by BO than the uniform grid search in the predefined target design space. We further provided the design variables of the selected optimal design for different porous materials formulations. Very interestingly, all the extracted optimal designs had more bacteria cells in the target growth region than pure biofilm growth in unconfined, substrate-less space. We conducted a comparison study to understand this phenomenon and found that there were 20\% more bacteria cells in the unconfined space than that confined in the porous materials. Furthermore, there were 128\% more bacteria cells in the target growth region for the porous substrate-induced biofilm growth compared with the unconfined space. We thence proposed that the existence of porous substrates stimulated the biofilms by taking up volumetric space to push growth upwards. Note that this is not universally tested for all kinds of porous materials with all radii range, and testing the side effects for confinement-induced biofilm growth would be our follow-up work in the future.

Our work is significant and innovative from three major aspects: (1) Implications and guidance to broad audiences. Our work could inspire theorists and programmers to develop new theories and algorithms for modeling biofilm and guide experimentalists to conduct new investigations. (2) Rigorous and comprehensive optimization analysis of the optimization process and direct characterization of the design space. (3) Understanding biophysical mechanisms from both the optimization characterization and computational modeling brings in new knowledge regarding the growth of biofilms. From these three aspects, our work bridges a broad range of different research areas spanning mechanics of materials, machine learning, and biology. To our knowledge, this is the first work that utilizes ML as an optimization tool for characterizing the underlying mechanisms of confined biofilm dynamics using computational models. We hope to inspire a new paradigm of conducting inverse design for physical discoveries by leveraging computational models, ML, and design optimizations.

\section*{Acknowledgement}

J.Y. acknowledges support from the US National Science Foundation under awards CMMI-2038057, ITE-2236190, and EFMA-2223785, as well as the Cornell University faculty startup grant. The authors also acknowledge the computational resources provided by the NSF Advanced Cyberinfrastructure Coordination Ecosystem: Services \& Support (ACCESS) program under grant BIO210063 and the computational resources provided by the G2 cluster from Cornell University.

\section*{Competing Interests}

The Authors declare no Competing Financial or Non-Financial Interests.

\section*{Data Availability}

The related data for optimization is available on \url{https://github.com/hanfengzhai/PyLAMDO}.

\section*{Author Contributions}

H.Z. and J.Y. designed and conceived the research. H.Z. wrote the code and performed the computations. H.Z. and J.Y. conducted data analysis. H.Z. and J.Y. wrote the manuscript. J.Y. acquired the funding and computational resources.

\section*{Appendix}

{\color{blue}Supplementary Figure 1} shows the whole optimization process updated by both EI and UCB acquisition functions for the porous membrane design case. The UCB exhibits more evident fluctuation along the sampling process and the EI acquisition sampled objectives are more ``clustered'' towards the upper region. To evaluate our findings more rigorously, {\color{blue}Supplementary Figure 1} \textbf{B} visualizes the overall statistical distribution of the objectives by two different acquisition functions. It can be qualitatively observed that the variance of EI is evidently smaller than that of UCB, and the mean objective value sampled by EI is higher. Quantitatively, the objective variance for the EI and UCB acquisition functions are $2.62\times10^{7}$ and $3.46\times10^{7}$, respectively, where the UCB acquisition sampled objectives' variance is 32.08\% higher than the EI acquisition. The mean objective values updated by EI and UCB acquisition functions are $\mathcal{N}_{\rm bio}^{\sf EI} = 30502$ \& $\mathcal{N}_{\rm bio}^{\sf UCB} = 30056$, respectively. The EI mean objective is 1.49\% higher than the UCB acquisition function. {\color{blue}Supplementary Figure 1} \textbf{C1} \& \textbf{C2} visualizes the trends of the normalized design variables along the sampling process by EI and UCB acquisition functions, respectively. It can be deduced from both the subfigures that $\bar{\mathcal{D}}$ is generally being sampled towards higher values and $N_{\rm unit}$ is being sampled in relatively lower values during the optimization processes, by observing their value range visualized by the color bar.

\begin{figure}[htbp]
    \centering
    \refstepcounter{figure}
    \renewcommand{\thefigure}{Supp. Fig. 1}
    \includegraphics[width=36em]{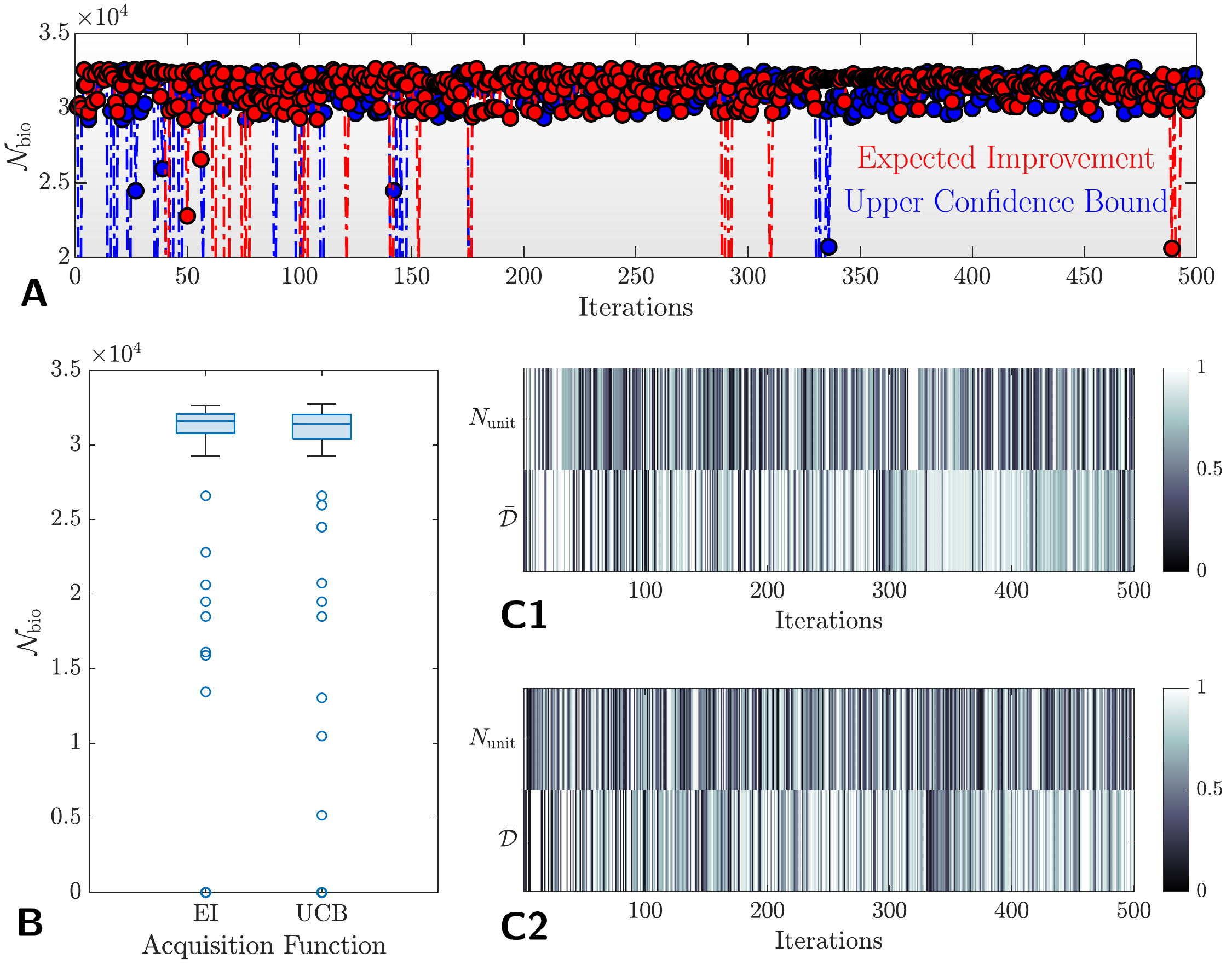}
    \caption*{{\color{blue}Supplementary Figure 1}: Optimization results for 2D porous membrane design. \textbf{(A)} The change of the objective value $\mathcal{N}_{\rm bio}$ along the iteration process. The red dotted dashed line represents the BO process through the Expected Improvement (EI) acquisition function. The blue dotted dashed line represents the BO process through the Upper Confidence Bound (UCB) acquisition function. \textbf{(B)} The statistical distribution of the objective values along the optimization processes characterized by the two different acquisition functions. \textbf{(C)} The normalized design variable change along the iteration process, corresponding to subfigure \textbf{A}, where subfigures \textbf{C1} \& \textbf{C2} represent the BO updated by EI and UCB acquisition function, respectively.}
    \label{fig3}
\end{figure}

{\color{blue}Supplementary Figure 2} visualizes the overall design processes for lattice metamaterials and 3D porous media, respectively (Supplementary Figure 2 \textbf{B} \& \textbf{C}). The upper figure (\textbf{1}) stands for the change of the objectives and the lower figure stand for the design variables' changes w.r.t. the iterations, similar to what has been shown in Figure \ref{fig3}. A converging process of the objective values is observed for 3D porous media (\textbf{B1}), whereas the objectives are most fluctuating more for the lattice metamaterials (\textbf{A1}), which can be attributed to the nonconvex design space in Figure \ref{fig6}. For the lattice metamaterials, the design variables are fluctuating along the iterations where $\bar{\mathcal{D}}$ is sampled toward higher values and $N_{\rm unit}$ is sampled toward the lower (\textbf{A2}). For the 3D porous media, similar trends are also observed yet the difference is they are initially sampled in a similar value range and the discrepancy of the sampling value trends begins to occur after approximately 300 iterations (\textbf{B2}).

\begin{figure}[htbp]
    \centering
    \refstepcounter{figure}
    \renewcommand{\thefigure}{Supp. Fig. 2}
    \includegraphics[width=45em]{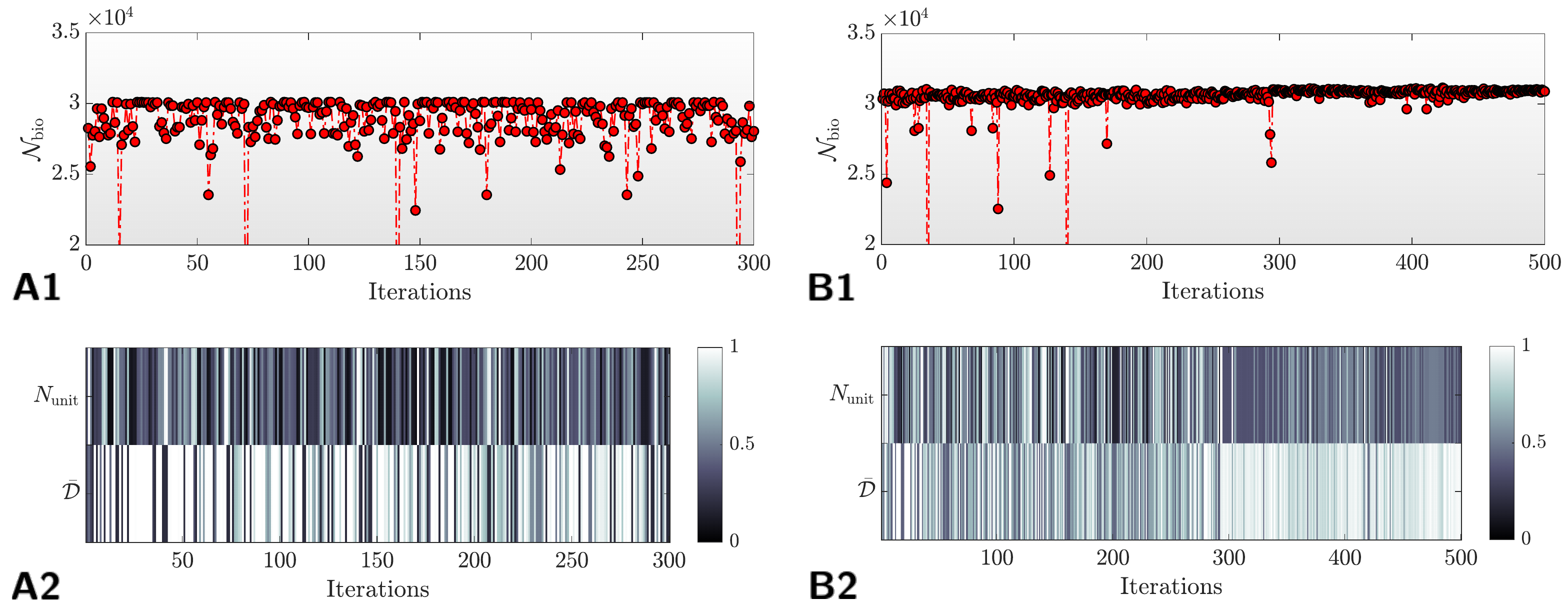}
    \caption*{{\color{blue}Supplementary Figure 2}: Design optimization process for both the lattice metamaterials and 3D porous media. The top subfigures indicate the objective value (bacteria cell numbers $\mathcal{N}_{\rm bio}$) along the simulation iteration processes. The bottom subfigures indicate the normalized design variable maps to visualize the changing trends of the design variables. \textbf{(A1)} Objective value change for the lattice metamaterials optimization case. \textbf{(A2)} Objective value change for the 3D porous media optimization case. \textbf{(B1)} The design variables' change for the lattice metamaterials optimization case. \textbf{(B2)} The design variables' change for the 3D porous media optimization case. }
    \label{suppfig1}
\end{figure}

\begin{figure}[htbp]
    \centering
    \refstepcounter{figure}
    \renewcommand{\thefigure}{Supp. Fig. 3}
    \includegraphics[width=35em]{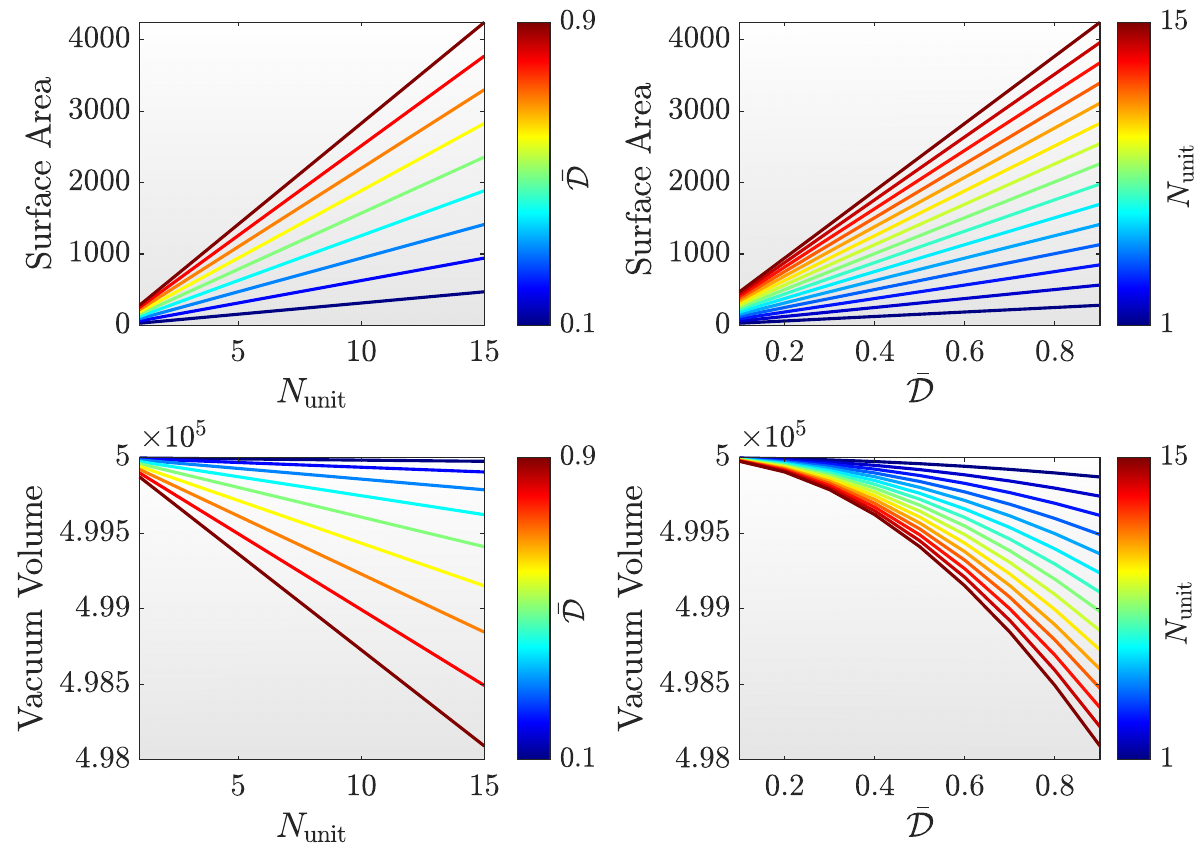}
    \caption*{{\color{blue}Supplementary Figure 3}: Model parametric relationships between the Surface Areas and Vacuum Volumes w.r.t. $N_{\rm unit}$ and $\bar{\mathcal{D}}$ (Equations \ref{opt_prob_equation}). The color gradients stand for the change of the other parameters different from the horizontal axis.}
    \label{suppfig2}
\end{figure}

{\color{blue}Supplementary Figure 3} visualizes the intrinsic relationship between the {\em Surface Areas} and {\em Vacuum Volumes} w.r.t. the design variables, $N_{\rm unit}$ and $\bar{\mathcal{D}}$. It can be generally concluded that the surface area varies linearly with both $N_{\rm unit}$ and $\bar{\mathcal{D}}$, but the vacuum volume is linear w.r.t. $N_{\rm unit}$ but nonlinear w.r.t. $\bar{\mathcal{D}}$. The goal of providing such a plot is to provide deeper insights into how the design variables reversibly affect the topological factors (i.e., surface areas and vacuum volumes), which mainly control the biofilm growth in the lattice scaffold case.

{\color{blue}Supplementary Table 1} shows the modeling parameters involved in the IbM models that are not completely listed in Section \ref{subsec_compmod}. Except for the growth rate $\xi_i$ are also given explicitly for algorithmic implementations, the IbM model we implemented also considered nutrient distribution as a form of linear-gradient, governed by the substrate and oxygen nutrient distribution factor, $\tt Diff_{sub}$ and $\tt Diff_{O_2}$, respectively\footnote{The detailed description of the algorithmic implementations of these parameters can be found in the Supplementary Information of Ref. \cite{Li2019}}. The kinetics parameter controlling the growth of {\em Heterotrophic Bacteria} is $\tt K_S^{HET}$. The yielding factor for both HET and EPS are $\tt Y_{HET}$ and $\tt Y_{EPS}$, respectively. In this model, the decay rates for both HET and EPS are set to zero.

\begin{table}[htbp]
    \centering
    \begin{tabular}{c c c c c}
        $\tt Diff_{sub}$ &  $\tt Diff_{O_2}$ & $\tt K_s^{HET}$ & $\tt Y_{HET}$ & $\tt Y_{EPS}$ \\  \hline \hspace{10pt}
        $1.6\times10^{-9}$ & $2.3\times10^{-9}$ & $3.5\times10^{-5}$ & 0.61 & 0.18
    \end{tabular}
    \caption*{{\color{blue}Supplementary Table 1}: The additional related parameters involved in the IbM simulation framework (in SI Unit).}
    \label{appendtab1}
\end{table}

\bibliographystyle{unsrtnat}
\bibliography{ref}

\begin{thebibliography}{52}
\providecommand{\natexlab}[1]{#1}
\providecommand{\url}[1]{\texttt{#1}}
\expandafter\ifx\csname urlstyle\endcsname\relax
  \providecommand{\doi}[1]{doi: #1}\else
  \providecommand{\doi}{doi: \begingroup \urlstyle{rm}\Url}\fi

\bibitem[Costerton et~al.(1999)Costerton, Stewart, and
  Greenberg]{biofilmmedref1}
J.~W. Costerton, Philip~S. Stewart, and E.~P. Greenberg.
\newblock Bacterial biofilms: A common cause of persistent infections.
\newblock \emph{Science}, 284\penalty0 (5418):\penalty0 1318--1322, May 1999.
\newblock \doi{10.1126/science.284.5418.1318}.
\newblock URL \url{https://doi.org/10.1126/science.284.5418.1318}.

\bibitem[Friedlander et~al.(2013)Friedlander, Vlamakis, Kim, Khan, Kolter, and
  Aizenberg]{pnas_activesurface}
Ronn~S. Friedlander, Hera Vlamakis, Philseok Kim, Mughees Khan, Roberto Kolter,
  and Joanna Aizenberg.
\newblock Bacterial flagella explore microscale hummocks and hollows to
  increase adhesion.
\newblock \emph{Proceedings of the National Academy of Sciences}, 110\penalty0
  (14):\penalty0 5624--5629, March 2013.
\newblock \doi{10.1073/pnas.1219662110}.
\newblock URL \url{https://doi.org/10.1073/pnas.1219662110}.

\bibitem[Feng et~al.(2015)Feng, Cheng, Wang, Borca-Tasciuc, Worobo, and
  Moraru]{surface_area_biofilm}
Guoping Feng, Yifan Cheng, Shu-Yi Wang, Diana~A Borca-Tasciuc, Randy~W Worobo,
  and Carmen~I Moraru.
\newblock Bacterial attachment and biofilm formation on surfaces are reduced by
  small-diameter nanoscale pores: how small is small enough?
\newblock \emph{npj Biofilms and Microbiomes}, 1\penalty0 (1), December 2015.
\newblock \doi{10.1038/npjbiofilms.2015.22}.
\newblock URL \url{https://doi.org/10.1038/npjbiofilms.2015.22}.

\bibitem[Yebra et~al.(2004)Yebra, Kiil, and Dam-Johansen]{marineref1}
Diego~Meseguer Yebra, S{\o}ren Kiil, and Kim Dam-Johansen.
\newblock Antifouling technology{\textemdash}past, present and future steps
  towards efficient and environmentally friendly antifouling coatings.
\newblock \emph{Progress in Organic Coatings}, 50\penalty0 (2):\penalty0
  75--104, July 2004.
\newblock \doi{10.1016/j.porgcoat.2003.06.001}.
\newblock URL \url{https://doi.org/10.1016/j.porgcoat.2003.06.001}.

\bibitem[Dobretsov et~al.(2006)Dobretsov, Dahms, and Qian]{marineref2}
Sergey Dobretsov, Hans-Uwe Dahms, and PerI-Yuan Qian.
\newblock Inhibition of biofouling by marine microorganisms and their
  metabolites.
\newblock \emph{Biofouling}, 22\penalty0 (1):\penalty0 43--54, January 2006.
\newblock \doi{10.1080/08927010500504784}.
\newblock URL \url{https://doi.org/10.1080/08927010500504784}.

\bibitem[Donlan and Costerton(2002)]{biofilmmedref2}
Rodney~M. Donlan and J.~William Costerton.
\newblock Biofilms: Survival mechanisms of clinically relevant microorganisms.
\newblock \emph{Clinical Microbiology Reviews}, 15\penalty0 (2):\penalty0
  167--193, April 2002.
\newblock \doi{10.1128/cmr.15.2.167-193.2002}.
\newblock URL \url{https://doi.org/10.1128/cmr.15.2.167-193.2002}.

\bibitem[Jonkers et~al.(2010)Jonkers, Thijssen, Muyzer, Copuroglu, and
  Schlangen]{concreteref1}
Henk~M. Jonkers, Arjan Thijssen, Gerard Muyzer, Oguzhan Copuroglu, and Erik
  Schlangen.
\newblock Application of bacteria as self-healing agent for the development of
  sustainable concrete.
\newblock \emph{Ecological Engineering}, 36\penalty0 (2):\penalty0 230--235,
  February 2010.
\newblock \doi{10.1016/j.ecoleng.2008.12.036}.
\newblock URL \url{https://doi.org/10.1016/j.ecoleng.2008.12.036}.

\bibitem[Chattopadhyay et~al.(2022)Chattopadhyay, J, Usman, and
  Varjani]{wastewatertreatref}
Indranil Chattopadhyay, Rajesh~Banu J, T.~M.~Mohamed Usman, and Sunita Varjani.
\newblock Exploring the role of microbial biofilm for industrial effluents
  treatment.
\newblock \emph{Bioengineered}, 13\penalty0 (3):\penalty0 6420--6440, February
  2022.
\newblock \doi{10.1080/21655979.2022.2044250}.
\newblock URL \url{https://doi.org/10.1080/21655979.2022.2044250}.

\bibitem[Balasubramanian et~al.(2019)Balasubramanian, Aubin-Tam, and
  Meyer]{3dbioprintingref}
Srikkanth Balasubramanian, Marie-Eve Aubin-Tam, and Anne~S. Meyer.
\newblock 3d printing for the fabrication of biofilm-based functional living
  materials.
\newblock \emph{{ACS} Synthetic Biology}, 8\penalty0 (7):\penalty0 1564--1567,
  July 2019.
\newblock \doi{10.1021/acssynbio.9b00192}.
\newblock URL \url{https://doi.org/10.1021/acssynbio.9b00192}.

\bibitem[Zhai and Yeo(2022)]{zhai_acs_biofilm}
Hanfeng Zhai and Jingjie Yeo.
\newblock Computational design of antimicrobial active surfaces via automated
  bayesian optimization.
\newblock \emph{ACS Biomaterials Science \& Engineering}, 9\penalty0
  (1):\penalty0 269–279, Dec 2022.
\newblock ISSN 2373-9878.
\newblock \doi{10.1021/acsbiomaterials.2c01079}.
\newblock URL \url{http://dx.doi.org/10.1021/acsbiomaterials.2c01079}.

\bibitem[Liu et~al.(2022)Liu, Ueki, Gao, Woodard, Nevin, Fu, Fu, Sun, Lovley,
  and Yao]{biofilm_clean_energy}
Xiaomeng Liu, Toshiyuki Ueki, Hongyan Gao, Trevor~L. Woodard, Kelly~P. Nevin,
  Tianda Fu, Shuai Fu, Lu~Sun, Derek~R. Lovley, and Jun Yao.
\newblock Microbial biofilms for electricity generation from water evaporation
  and power to wearables.
\newblock \emph{Nature Communications}, 13\penalty0 (1), July 2022.
\newblock \doi{10.1038/s41467-022-32105-6}.
\newblock URL \url{https://doi.org/10.1038/s41467-022-32105-6}.

\bibitem[Mukherjee and Cao(2020)]{biofilm_control_ref}
Manisha Mukherjee and Bin Cao.
\newblock Engineering controllable biofilms for biotechnological applications.
\newblock \emph{Microbial Biotechnology}, 14\penalty0 (1):\penalty0 74--78,
  November 2020.
\newblock \doi{10.1111/1751-7915.13715}.
\newblock URL \url{https://doi.org/10.1111/1751-7915.13715}.

\bibitem[Powell et~al.(2018)Powell, Pritchard, Ferguson, Powell, Patel, Rye,
  Sakellakou, Buurma, Brilliant, Copping, Menzies, Lewis, Hill, and
  Thomas]{md_biofilm_study}
Lydia~C. Powell, Manon~F. Pritchard, Elaine~L. Ferguson, Kate~A. Powell,
  Shree~U. Patel, Phil~D. Rye, Stavroula-Melina Sakellakou, Niklaas~J. Buurma,
  Charles~D. Brilliant, Jack~M. Copping, Georgina~E. Menzies, Paul~D. Lewis,
  Katja~E. Hill, and David~W. Thomas.
\newblock Targeted disruption of the extracellular polymeric network of
  pseudomonas aeruginosa biofilms by alginate oligosaccharides.
\newblock \emph{npj Biofilms and Microbiomes}, 4\penalty0 (1), June 2018.
\newblock \doi{10.1038/s41522-018-0056-3}.
\newblock URL \url{https://doi.org/10.1038/s41522-018-0056-3}.

\bibitem[Xu et~al.(2011)Xu, Meakin, Tartakovsky, and
  Scheibe]{dpd_biofilm_model}
Zhijie Xu, Paul Meakin, Alexandre Tartakovsky, and Timothy~D. Scheibe.
\newblock Dissipative-particle-dynamics model of biofilm growth.
\newblock \emph{Physical Review E}, 83\penalty0 (6), June 2011.
\newblock \doi{10.1103/physreve.83.066702}.
\newblock URL \url{https://doi.org/10.1103/physreve.83.066702}.

\bibitem[Brandani et~al.(2015)Brandani, Schor, Morris, Stanley-Wall, MacPhee,
  Marenduzzo, and Zachariae]{cgmd_biofilm}
Giovanni~B. Brandani, Marieke Schor, Ryan Morris, Nicola Stanley-Wall, Cait~E.
  MacPhee, Davide Marenduzzo, and Ulrich Zachariae.
\newblock The bacterial hydrophobin {BslA} is a switchable ellipsoidal janus
  nanocolloid.
\newblock \emph{Langmuir}, 31\penalty0 (42):\penalty0 11558--11563, October
  2015.
\newblock \doi{10.1021/acs.langmuir.5b02347}.
\newblock URL \url{https://doi.org/10.1021/acs.langmuir.5b02347}.

\bibitem[Smith et~al.(2007)Smith, Vaughan, and Chopp]{biofilm_fem_model}
Bryan Smith, Benjamin Vaughan, and David Chopp.
\newblock The extended finite element method for boundary layer problems in
  biofilm growth.
\newblock \emph{Communications in Applied Mathematics and Computational
  Science}, 2\penalty0 (1):\penalty0 35--56, August 2007.
\newblock \doi{10.2140/camcos.2007.2.35}.
\newblock URL \url{https://doi.org/10.2140/camcos.2007.2.35}.

\bibitem[Li et~al.(2019)Li, Taniguchi, Gedara, Gogulancea, Gonzalez-Cabaleiro,
  Chen, McGough, Ofiteru, Curtis, and Zuliani]{Li2019}
Bowen Li, Denis Taniguchi, Jayathilake~Pahala Gedara, Valentina Gogulancea,
  Rebeca Gonzalez-Cabaleiro, Jinju Chen, Andrew~Stephen McGough, Irina~Dana
  Ofiteru, Thomas~P. Curtis, and Paolo Zuliani.
\newblock {NUFEB}: A massively parallel simulator for individual-based
  modelling of microbial communities.
\newblock \emph{{PLOS} Computational Biology}, 15\penalty0 (12):\penalty0
  e1007125, December 2019.
\newblock \doi{10.1371/journal.pcbi.1007125}.
\newblock URL \url{https://doi.org/10.1371/journal.pcbi.1007125}.

\bibitem[Kapellos et~al.(2015)Kapellos, Alexiou, and
  Pavlou]{biofilm_porescale_mircometer}
George~E. Kapellos, Terpsichori~S. Alexiou, and Stavros Pavlou.
\newblock Fluid-biofilm interactions in porous media.
\newblock In \emph{Heat Transfer and Fluid Flow in Biological Processes}, pages
  207--238. Elsevier, 2015.
\newblock \doi{10.1016/b978-0-12-408077-5.00008-0}.
\newblock URL \url{https://doi.org/10.1016/b978-0-12-408077-5.00008-0}.

\bibitem[Galy et~al.(2012)Galy, Latour-Lambert, Zrelli, Ghigo, Beloin, and
  Henry]{biofilm_mechanical_adhesion_micrometer}
Olivier Galy, Patricia Latour-Lambert, Kais Zrelli, Jean-Marc Ghigo, Christophe
  Beloin, and Nelly Henry.
\newblock Mapping of bacterial biofilm local mechanics by magnetic
  microparticle actuation.
\newblock \emph{Biophysical Journal}, 103\penalty0 (6):\penalty0 1400--1408,
  September 2012.
\newblock \doi{10.1016/j.bpj.2012.07.001}.
\newblock URL \url{https://doi.org/10.1016/j.bpj.2012.07.001}.

\bibitem[Rodrigo-Navarro et~al.(2021)Rodrigo-Navarro, Sankaran, Dalby, del
  Campo, and Salmeron-Sanchez]{englivmat_natrevmat_reviewpaper}
Aleixandre Rodrigo-Navarro, Shrikrishnan Sankaran, Matthew~J. Dalby,
  Ar{\'{a}}nzazu del Campo, and Manuel Salmeron-Sanchez.
\newblock Engineered living biomaterials.
\newblock \emph{Nature Reviews Materials}, 6\penalty0 (12):\penalty0
  1175--1190, August 2021.
\newblock \doi{10.1038/s41578-021-00350-8}.
\newblock URL \url{https://doi.org/10.1038/s41578-021-00350-8}.

\bibitem[Li et~al.(2021)Li, Xie, and Yingling]{md_fulltime_6months}
Nan~K. Li, Yuxin Xie, and Yaroslava~G. Yingling.
\newblock Insights into structure and aggregation behavior of elastin-like
  polypeptide coacervates: All-atom molecular dynamics simulations.
\newblock \emph{The Journal of Physical Chemistry B}, 125\penalty0
  (30):\penalty0 8627--8635, July 2021.
\newblock \doi{10.1021/acs.jpcb.1c02822}.
\newblock URL \url{https://doi.org/10.1021/acs.jpcb.1c02822}.

\bibitem[Duddu et~al.(2008)Duddu, Bordas, Chopp, and
  Moran]{biofilm_xfem_levelset_ijnme}
Ravindra Duddu, St{\'{e}}phane Bordas, David Chopp, and Brian Moran.
\newblock A combined extended finite element and level set method for biofilm
  growth.
\newblock \emph{International Journal for Numerical Methods in Engineering},
  74\penalty0 (5):\penalty0 848--870, 2008.
\newblock \doi{10.1002/nme.2200}.
\newblock URL \url{https://doi.org/10.1002/nme.2200}.

\bibitem[Zhai et~al.(2022)Zhai, Zhou, and Hu]{bubblenet}
Hanfeng Zhai, Quan Zhou, and Guohui Hu.
\newblock {Predicting micro-bubble dynamics with semi-physics-informed deep
  learning}.
\newblock \emph{AIP Advances}, 12\penalty0 (3), 03 2022.
\newblock ISSN 2158-3226.
\newblock \doi{10.1063/5.0079602}.
\newblock URL \url{https://doi.org/10.1063/5.0079602}.
\newblock 035153.

\bibitem[Hadamard(1902)]{hadamard_illposed}
J.~Hadamard.
\newblock On problems with partial derivatives and their physical significance.
\newblock \emph{Princeton University Bulletin}, 13:\penalty0 49--52, 1902.

\bibitem[Frazier(2018)]{bayes}
Peter~I. Frazier.
\newblock A tutorial on bayesian optimization, 2018.
\newblock URL \url{https://arxiv.org/abs/1807.02811}.

\bibitem[Fuhg and Bouklas(2022)]{Jan_PIGPR}
Jan~N. Fuhg and Nikolaos Bouklas.
\newblock On physics-informed data-driven isotropic and anisotropic
  constitutive models through probabilistic machine learning and space-filling
  sampling.
\newblock \emph{Computer Methods in Applied Mechanics and Engineering},
  394:\penalty0 114915, May 2022.
\newblock \doi{10.1016/j.cma.2022.114915}.
\newblock URL \url{https://doi.org/10.1016/j.cma.2022.114915}.

\bibitem[Sutton and Barto(2018)]{drl_intro_book_ref}
Richard~S Sutton and Andrew~G Barto.
\newblock \emph{Reinforcement Learning: An Introduction}.
\newblock MIT Press, 2018.
\newblock URL
  \url{https://web.stanford.edu/class/psych209/Readings/SuttonBartoIPRLBook2ndEd.pdf}.

\bibitem[Mitchell(1998)]{ga_paper_ref}
Melanie Mitchell.
\newblock \emph{An Introduction to Genetic Algorithms}.
\newblock MIT Press, 1998.

\bibitem[Kennedy and Eberhart(1995)]{pso_paper_ref}
J.~Kennedy and R.~Eberhart.
\newblock Particle swarm optimization.
\newblock In \emph{Proceedings of ICNN'95 - International Conference on Neural
  Networks}, volume~4, pages 1942--1948 vol.4, 1995.
\newblock \doi{10.1109/ICNN.1995.488968}.

\bibitem[Monod(1949)]{Monod1949}
Jacques Monod.
\newblock {THE} {GROWTH} {OF} {BACTERIAL} {CULTURES}.
\newblock \emph{Annual Review of Microbiology}, 3\penalty0 (1):\penalty0
  371--394, October 1949.
\newblock \doi{10.1146/annurev.mi.03.100149.002103}.
\newblock URL \url{https://doi.org/10.1146/annurev.mi.03.100149.002103}.

\bibitem[Xavier et~al.(2005)Xavier, Picioreanu, and van
  Loosdrecht]{eps_original_formulation}
Joao~B. Xavier, Cristian Picioreanu, and Mark C.~M. van Loosdrecht.
\newblock A framework for multidimensional modelling of activity and structure
  of multispecies biofilms.
\newblock \emph{Environmental Microbiology}, 7\penalty0 (8):\penalty0
  1085--1103, August 2005.
\newblock \doi{10.1111/j.1462-2920.2005.00787.x}.
\newblock URL \url{https://doi.org/10.1111/j.1462-2920.2005.00787.x}.

\bibitem[Jayathilake et~al.(2017)Jayathilake, Gupta, Li, Madsen, Oyebamiji,
  Gonz{\'{a}}lez-Cabaleiro, Rushton, Bridgens, Swailes, Allen, McGough,
  Zuliani, Ofiteru, Wilkinson, Chen, and Curtis]{nufeb_eps_formulation}
Pahala~Gedara Jayathilake, Prashant Gupta, Bowen Li, Curtis Madsen, Oluwole
  Oyebamiji, Rebeca Gonz{\'{a}}lez-Cabaleiro, Steve Rushton, Ben Bridgens,
  David Swailes, Ben Allen, A.~Stephen McGough, Paolo Zuliani, Irina~Dana
  Ofiteru, Darren Wilkinson, Jinju Chen, and Tom Curtis.
\newblock A mechanistic individual-based model of microbial communities.
\newblock \emph{{PLOS} {ONE}}, 12\penalty0 (8):\penalty0 e0181965, August 2017.
\newblock \doi{10.1371/journal.pone.0181965}.
\newblock URL \url{https://doi.org/10.1371/journal.pone.0181965}.

\bibitem[Rasmussen and Williams(2006)]{gp_book}
Carl~Edward Rasmussen and Christopher K.~I. Williams.
\newblock \emph{Gaussian Processes for Machine Learning}.
\newblock The MIT Press, 2006.
\newblock ISBN 0-262-18253-X.

\bibitem[Deshwal et~al.(2021)Deshwal, Simon, and
  Doppa]{msde_nanoporous_material}
Aryan Deshwal, Cory~M. Simon, and Janardhan~Rao Doppa.
\newblock Bayesian optimization of nanoporous materials.
\newblock \emph{Molecular Systems Design {\&} Engineering}, 6\penalty0
  (12):\penalty0 1066--1086, 2021.
\newblock \doi{10.1039/d1me00093d}.
\newblock URL \url{https://doi.org/10.1039/d1me00093d}.

\bibitem[Snoek et~al.(2012)Snoek, Larochelle, and Adams]{acquisition_func}
Jasper Snoek, Hugo Larochelle, and Ryan~P. Adams.
\newblock Practical bayesian optimization of machine learning algorithms, 2012.
\newblock URL \url{https://arxiv.org/abs/1206.2944}.

\bibitem[Pousti et~al.(2019)Pousti, Zarabadi, Amirdehi, Paquet-Mercier, and
  Greener]{biofilm_chanel_flow_review}
Mohammad Pousti, Mir~Pouyan Zarabadi, Mehran~Abbaszadeh Amirdehi,
  Fran{\c{c}}ois Paquet-Mercier, and Jesse Greener.
\newblock Microfluidic bioanalytical flow cells for biofilm studies: a review.
\newblock \emph{The Analyst}, 144\penalty0 (1):\penalty0 68--86, 2019.
\newblock \doi{10.1039/c8an01526k}.
\newblock URL \url{https://doi.org/10.1039/c8an01526k}.

\bibitem[Ye et~al.(2021)Ye, Zhang, Li, Zhu, Chen, and
  Liao]{channel_flow_anode_energy}
Dingding Ye, Pengqing Zhang, Jun Li, Xun Zhu, Rong Chen, and Qiang Liao.
\newblock In situ visualization of biofilm formation in a microchannel for a
  microfluidic microbial fuel cell anode.
\newblock \emph{International Journal of Hydrogen Energy}, 46\penalty0
  (27):\penalty0 14651--14658, April 2021.
\newblock \doi{10.1016/j.ijhydene.2020.08.170}.
\newblock URL \url{https://doi.org/10.1016/j.ijhydene.2020.08.170}.

\bibitem[Landa-Marb{\'{a}}n et~al.(2019)Landa-Marb{\'{a}}n, Pop, Kumar, and
  Radu]{channel_flow_numerical_study}
David Landa-Marb{\'{a}}n, Iuliu~Sorin Pop, Kundan Kumar, and Florin~A. Radu.
\newblock Numerical simulation of biofilm formation in a microchannel.
\newblock In \emph{Lecture Notes in Computational Science and Engineering},
  pages 799--807. Springer International Publishing, 2019.
\newblock \doi{10.1007/978-3-319-96415-7_75}.
\newblock URL \url{https://doi.org/10.1007/978-3-319-96415-7_75}.

\bibitem[Aspa et~al.(2011)Aspa, Debenest, and
  Quintard]{channel_flow_numerical_study_2}
Y.~Aspa, G.~Debenest, and M.~Quintard.
\newblock Effective dispersion in channelled biofilms.
\newblock \emph{International Journal of Environment and Waste Management},
  7\penalty0 (1/2):\penalty0 112, 2011.
\newblock \doi{10.1504/ijewm.2011.037371}.
\newblock URL \url{https://doi.org/10.1504/ijewm.2011.037371}.

\bibitem[Landa-Marb{\'{a}}n et~al.(2020)Landa-Marb{\'{a}}n, B{\o}dtker, Kumar,
  Pop, and Radu]{biofilm_chanel_theoretical_ref}
David Landa-Marb{\'{a}}n, Gunhild B{\o}dtker, Kundan Kumar, Iuliu~S. Pop, and
  Florin~A. Radu.
\newblock An upscaled model for permeable biofilm in a thin channel and tube.
\newblock \emph{Transport in Porous Media}, 132\penalty0 (1):\penalty0 83--112,
  January 2020.
\newblock \doi{10.1007/s11242-020-01381-5}.
\newblock URL \url{https://doi.org/10.1007/s11242-020-01381-5}.

\bibitem[Ma et~al.(2022)Ma, Chang, Wu, and Zhao]{renee_dl_design}
Chunping Ma, Yilong Chang, Shuai Wu, and Ruike~Renee Zhao.
\newblock Deep learning-accelerated designs of tunable magneto-mechanical
  metamaterials.
\newblock \emph{{ACS} Applied Materials {\&} Interfaces}, 14\penalty0
  (29):\penalty0 33892--33902, July 2022.
\newblock \doi{10.1021/acsami.2c09052}.
\newblock URL \url{https://doi.org/10.1021/acsami.2c09052}.

\bibitem[Shaw et~al.(2019)Shaw, Sun, Portela, Barranco, Greer, and
  Hopkins]{carlos_natcomm_design}
Lucas~A. Shaw, Frederick Sun, Carlos~M. Portela, Rodolfo~I. Barranco, Julia~R.
  Greer, and Jonathan~B. Hopkins.
\newblock Computationally efficient design of directionally compliant
  metamaterials.
\newblock \emph{Nature Communications}, 10\penalty0 (1), January 2019.
\newblock \doi{10.1038/s41467-018-08049-1}.
\newblock URL \url{https://doi.org/10.1038/s41467-018-08049-1}.

\bibitem[Gu(2018)]{wendy_nano_architectured}
X.~Wendy Gu.
\newblock Mechanical properties of architected nanomaterials made from
  organic{\textendash}inorganic nanocrystals.
\newblock \emph{{JOM}}, 70\penalty0 (10):\penalty0 2205--2217, August 2018.
\newblock \doi{10.1007/s11837-018-3094-7}.
\newblock URL \url{https://doi.org/10.1007/s11837-018-3094-7}.

\bibitem[Portela et~al.(2020)Portela, Vidyasagar, Kr\"{o}del, Weissenbach, Yee,
  Greer, and Kochmann]{carlos_pnas}
Carlos~M. Portela, A.~Vidyasagar, Sebastian Kr\"{o}del, Tamara Weissenbach,
  Daryl~W. Yee, Julia~R. Greer, and Dennis~M. Kochmann.
\newblock Extreme mechanical resilience of self-assembled nanolabyrinthine
  materials.
\newblock \emph{Proceedings of the National Academy of Sciences}, 117\penalty0
  (11):\penalty0 5686--5693, March 2020.
\newblock \doi{10.1073/pnas.1916817117}.
\newblock URL \url{https://doi.org/10.1073/pnas.1916817117}.

\bibitem[Ovelheiro(2020)]{lattice_mat_biofilm_ref_MSThesis}
Bryan Ovelheiro.
\newblock 3d printed architected materials for improving biofilm carriers for
  wastewater treatment applications.
\newblock Masters project, University of Massachusetts Amherst, Amherst, MA,
  2020.
\newblock URL \url{https://scholarworks.umass.edu/cee_masters/20/}.

\bibitem[He et~al.(2021)He, Fu, Pang, Li, Li, Zhu, Lu, Sun, Liao, and
  Schr\"{o}der]{lattice_mat_biofilm_ref_energy}
Yu-Ting He, Qian Fu, Yuan Pang, Qing Li, Jun Li, Xun Zhu, Ren-Hao Lu, Wei Sun,
  Qiang Liao, and Uwe Schr\"{o}der.
\newblock Customizable design strategies for high-performance bioanodes in
  bioelectrochemical systems.
\newblock \emph{{iScience}}, 24\penalty0 (3):\penalty0 102163, March 2021.
\newblock \doi{10.1016/j.isci.2021.102163}.
\newblock URL \url{https://doi.org/10.1016/j.isci.2021.102163}.

\bibitem[Hall et~al.(2021)Hall, Palmer, Ji, Ehrlich, and
  Kr{\'{o}}l]{lattice_mat_biofilm_ref_biomed}
Donald~C. Hall, Phillip Palmer, Hai-Feng Ji, Garth~D. Ehrlich, and
  Jaros{\l}aw~E. Kr{\'{o}}l.
\newblock Bacterial biofilm growth on 3d-printed materials.
\newblock \emph{Frontiers in Microbiology}, 12, May 2021.
\newblock \doi{10.3389/fmicb.2021.646303}.
\newblock URL \url{https://doi.org/10.3389/fmicb.2021.646303}.

\bibitem[Bhattacharjee and Datta(2019)]{natcomm_bacteria_hopping_porous}
Tapomoy Bhattacharjee and Sujit~S. Datta.
\newblock Bacterial hopping and trapping in porous media.
\newblock \emph{Nature Communications}, 10\penalty0 (1), May 2019.
\newblock \doi{10.1038/s41467-019-10115-1}.
\newblock URL \url{https://doi.org/10.1038/s41467-019-10115-1}.

\bibitem[Carrel et~al.(2018)Carrel, Morales, Beltran, Derlon, Kaufmann,
  Morgenroth, and Holzner]{biofim_porous_3D_ref}
Maxence Carrel, Ver{\'{o}}nica~L. Morales, Mario~A. Beltran, Nicolas Derlon,
  Rolf Kaufmann, Eberhard Morgenroth, and Markus Holzner.
\newblock Biofilms in 3d porous media: Delineating the influence of the pore
  network geometry, flow and mass transfer on biofilm development.
\newblock \emph{Water Research}, 134:\penalty0 280--291, May 2018.
\newblock \doi{10.1016/j.watres.2018.01.059}.
\newblock URL \url{https://doi.org/10.1016/j.watres.2018.01.059}.

\bibitem[Coyte et~al.(2016)Coyte, Tabuteau, Gaffney, Foster, and
  Durham]{pnas_biofilm_competetion}
Katharine~Z. Coyte, Herv{\'{e}} Tabuteau, Eamonn~A. Gaffney, Kevin~R. Foster,
  and William~M. Durham.
\newblock Microbial competition in porous environments can select against rapid
  biofilm growth.
\newblock \emph{Proceedings of the National Academy of Sciences}, 114\penalty0
  (2), December 2016.
\newblock \doi{10.1073/pnas.1525228113}.
\newblock URL \url{https://doi.org/10.1073/pnas.1525228113}.

\bibitem[Kurz et~al.(2022)Kurz, Secchi, Carrillo, Bourg, Stocker, and
  Jimenez-Martinez]{pnas_porous_flow}
Dorothee~L. Kurz, Eleonora Secchi, Francisco~J. Carrillo, Ian~C. Bourg, Roman
  Stocker, and Joaquin Jimenez-Martinez.
\newblock Competition between growth and shear stress drives intermittency in
  preferential flow paths in porous medium biofilms.
\newblock \emph{Proceedings of the National Academy of Sciences}, 119\penalty0
  (30), July 2022.
\newblock \doi{10.1073/pnas.2122202119}.
\newblock URL \url{https://doi.org/10.1073/pnas.2122202119}.

\bibitem[Dehkharghani et~al.(2023)Dehkharghani, Waisbord, and
  Guasto]{commphys_scale_pore_effect}
Amin Dehkharghani, Nicolas Waisbord, and Jeffrey~S. Guasto.
\newblock Self-transport of swimming bacteria is impaired by porous
  microstructure.
\newblock \emph{Communications Physics}, 6\penalty0 (1), January 2023.
\newblock \doi{10.1038/s42005-023-01136-w}.
\newblock URL \url{https://doi.org/10.1038/s42005-023-01136-w}.

\end{thebibliography}

\end{document}